\documentclass[twocolumn]{aastex631}

\usepackage{my_astro}
\usepackage{courier}
\usepackage{amsmath}
\usepackage{nth}
\usepackage[all]{hypcap} % this makes clicking a figure link take you to the figure instead of the caption
\usepackage[normalem]{ulem} % for striking out text
\usepackage{aas_macros}

\newcommand{\mMcme}{m_\mathrm{CME}}
\newcommand{\Mdot}{$\dot{M}$}
\newcommand{\Mdotsun}{$\dot{M}_\odot$}

\newcommand{\mMdot}{\dot{M}}

\newcommand{\mMref}{m_\mathrm{ref}}

\newcommand{\dmax}{$\delta_\mathrm{max}$}
\newcommand{\mdmax}{\delta_\mathrm{max}}
\newcommand{\Fpre}{$F_\mathrm{pre}$}
\newcommand{\mFpre}{F_\mathrm{pre}}
\newcommand{\ee}{$\mathrm{\epsilon}$~Eri}
\newcommand{\linedesc}{Vertical gray lines are the \nth{18}, \nth{50}, and \nth{84} percentiles.}

\newcommand{\fitunits}{Symbols are the pre-flare flux, \Fpre~($10^{-15}$~\fluxcgs); CME acceleration, a (m~s$^{-2}$); dimming depth, \dmax (relative units); and the white noise hyperparameter, $\sigma$ (relative units).}

\renewcommand{\edit}[1]{}

\begin{document}

\title{Constraining the Physical Properties of Stellar Coronal Mass Ejections with Coronal Dimming: Application to Far Ultraviolet Data of $\epsilon$ Eridani}

% arXiv title
% Constraining the Physical Properties of Stellar Coronal Mass Ejections with Coronal Dimming: Application to Far Ultraviolet Data of $\epsilon$ Eridani

\correspondingauthor{R. O. Parke Loyd}
\email{astroparke@gmail.com}

\author{R. O. Parke Loyd}
\affiliation{Eureka Scientific, 2452 Delmer Street Suite 100, Oakland, CA, 94602-3017}
\affiliation{School of Earth and Space Exploration, Arizona State University, 781 E. Terrace Mall, Tempe, AZ 85287, USA}

\author{James Mason}
\affiliation{Applied Physics Laboratory, Johns Hopkins University, 11100 Johns Hopkins Road, Laurel, MD 20723, USA}

\author{Meng Jin}
\affiliation{SETI Institute, 189 N Bernardo Ave suite 200, Mountain View, CA 94043, USA}

\author{Evgenya L. Shkolnik}
\affiliation{School of Earth and Space Exploration, Arizona State University, 781 E. Terrace Mall, Tempe, AZ 85287, USA}

\author{Kevin France}
\affiliation{Laboratory for Atmospheric and Space Physics, University of Colorado, 600 UCB, Boulder, CO 80309, USA}

\author{Allison Youngblood}
\affiliation{Exoplanets and Stellar Astrophysics Laboratory, NASA Goddard Space Flight Center, Greenbelt, MD 20771, USA}

\author{Jackie Villadsen}
\affiliation{National Radio Astronomy Observatory, 520 Edgemont Rd., Charlottesville, VA 22903, USA}

\author{Christian Schneider}
\affiliation{Hamburger Sternwarte, Universität Hamburg, Gojenbergsweg 112, D-21029 Hamburg, Germany}

\author{Adam C. Schneider}
\affiliation{United States Naval Observatory, Flagstaff Station, 10391 West Naval Observatory Rd., Flagstaff, AZ 86005, USA}
\affiliation{Department of Physics and Astronomy, George Mason University, MS3F3, 4400 University Drive, Fairfax, VA 22030, USA}

\author{Joseph Llama}
\affiliation{Lowell Observatory, 1400 W Mars Hill Rd, Flagstaff, AZ 86001, USA}

\author{Tahina Ramiaramanantsoa}
\affiliation{School of Earth and Space Exploration, Arizona State University, 781 E. Terrace Mall, Tempe, AZ 85287, USA}

\author{Tyler Richey-Yowell}
\affiliation{School of Earth and Space Exploration, Arizona State University, 781 E. Terrace Mall, Tempe, AZ 85287, USA}

% arXiv authors
% R. O. Parke Loyd, James Mason, Meng Jin, Evgenya L. Shkolnik, Kevin France, Allison Youngblood, Jackie Villadsen, Christian Schneider, Adam C. Schneider, Joseph Llama, Tahina Ramiaramanantsoa, Tyler Richey-Yowell

\begin{abstract}
Coronal mass ejections (CMEs) are a prominent contributor to solar system space weather and might have impacted the Sun's early angular momentum evolution.
A signal diagnostic of CMEs on the Sun is coronal dimming: a drop in coronal emission, tied to the mass of the CME, that is the direct result of removing emitting plasma from the corona.
We present the results of a coronal dimming analysis of \Fexii~1349~\AA\ and \Fexxi~1354~\AA\ emission from \ee, a young K2 dwarf, with archival far-ultraviolet observations by the Hubble Space Telescope's Cosmic Origins Spectrograph.
Following a flare in February 2015, \ee's \Fexxi\ emission declined by $81\pm5$\%.
% chain = db.load_pickle(rc.path_fit_chain('a', 'Fe21'))
% chain = chain['chain']
% utils.mcmc_errors(chain[:,2])
Although enticing, a scant 3.8~min of preflare observations allows for the possibility that the \Fexxi\ decline was the decay of an earlier, unseen flare.
Dimming nondetections following each of three prominent flares constrain the possible mass of ejected \Fexii-emitting (1~MK) plasma to less than a few $\times10^{15}$~g.
This implies that CMEs ejecting this much or more 1~MK plasma occur less than a few times per day on \ee.
On the Sun, $10^{15}$~g CMEs occur once every few days.
For \ee, the mass loss rate due to CME-ejected 1~MK plasma could be $<0.6$~\Mdotsun, well below the star's \edit1{estimated} 30~\Mdotsun\ mass loss rate (wind + CMEs).
The order-of-magnitude formalism we developed for these mass estimates can be broadly applied to coronal dimming observations of any star.

\end{abstract}

% arXiv abstract
% Coronal mass ejections (CMEs) are a prominent contributor to solar system space weather and might have impacted the Sun's early angular momentum evolution. A signal diagnostic of CMEs on the Sun is coronal dimming: a drop in coronal emission, tied to the mass of the CME, that is the direct result of removing emitting plasma from the corona. We present the results of a coronal dimming analysis of Fe XII 1349 A and Fe XXI 1354 A emission from $\epsilon$ Eridani ($\epsilon$ Eri), a young K2 dwarf, with archival far-ultraviolet observations by the Hubble Space Telescope's Cosmic Origins Spectrograph. Following a flare in February 2015, $\epsilon$ Eri's Fe XXI emission declined by $81\pm5$%. Although enticing, a scant 3.8 min of preflare observations allows for the possibility that the Fe XXI decline was the decay of an earlier, unseen flare. Dimming nondetections following each of three prominent flares constrain the possible mass of ejected Fe XII-emitting (1 MK) plasma to less than a few $\times10^{15}$ g. This implies that CMEs ejecting this much or more 1 MK plasma occur less than a few times per day on $\epsilon$ Eri. On the Sun, $10^{15}$ g CMEs occur once every few days. For $\epsilon$ Eri, the mass loss rate due to CME-ejected 1 MK plasma could be $<0.6$ $\dot{M}_\odot$, well below the star's estimated 30 $\dot{M}_\odot$ mass loss rate (wind + CMEs). The order-of-magnitude formalism we developed for these mass estimates can be broadly applied to coronal dimming observations of any star.
\section{Introduction}

% general, stellar evolution, disks, planets, sun observations, dimming

Coronal mass ejections (CMEs) are a prominent feature of the dynamic space environment of the Sun.
When CMEs and associated particle showers impact planets, they disturb atmospheric chemistry, trigger magnetic storms, and strip atmospheric mass \citep{howard14,jakosky15}.

Stars with Sun-like magnetic dynamos likely also produce CMEs. CMEs are products of magnetic reconnection, and magnetic reconnection events that produce stellar flares are ubiquitous among FGKM stars \citep[e.g., ][]{davenport16}.
On the Sun, more energetic X-ray flares correlate with more massive CMEs \citep{aarnio11}.
\edit1{Applying solar relationships between CME mass and flare energy to the more frequent flares of active stars yields large rates of mass, energy, and angular momentum loss \citep{aarnio12,drake13,osten15}.
Predicted energy loss can reach an implausible fraction of the stellar luminosity and predicted CME masses can exceed a reasonable estimate for a star's total coronal mass, suggesting that such scalings break down in the regime of high stellar activity \citep{drake13}.
Predicted CME mass loss exceeds an observational estimate of the stellar wind for the young solar analog $\mathrm{pi}^1$~UMa by over two orders of magnitude \citep{wood14,drake16}.
Nonetheless, these scalings hint that the CME activity of young, active stars could be important to the evolution of the star, its disk, and its planets.}

Finding unambiguous evidence of stellar CMEs has proven challenging.
Searches for Type II radio bursts (emission from plasma oscillations that declines in frequency as the CME expands) have come up empty handed \citep[e.g., ][]{villadsen19}.
Doppler shifts in emission lines (namely H$\mathrm{\alpha}$) due to ejected plasma or drops in broadband X-ray emission attributed to absorption from a CME ejected into the line of sight have yielded some evidence of CMEs.
However, such events can be explained as chromospheric evaporation and brightening and changes in the coronal emission measure distribution \citep{argiroffi19}.
\edit1{Recently, \cite{namekata21} identified a prominence eruption (commonly associated with a CME) via blueshifted \Ha\ absorption following a superflare on the young solar-type star EK~Dra.}

\edit1{The scarcity and ambiguity of CME detections on active, main-sequence stars could indicate these stars' strong global magnetic fields trap CMEs \citep{drake13,drake16,alvarado18,alvarado19,alvarado20} or even inhibit their formation \citep{sun22}.
CMEs that do break free could be sapped of substantial energy, a picture consistent with the properties of existing CME candidates \citep{moschou19,namekata21}.
To fully understand the scarcity of stellar CME detections and the associated impact of CMEs on stellar and planetary evolution, our community must develop techniques to either reliably detect these events or prove their absence.}

\begin{figure*}
\includegraphics[width=\textwidth]{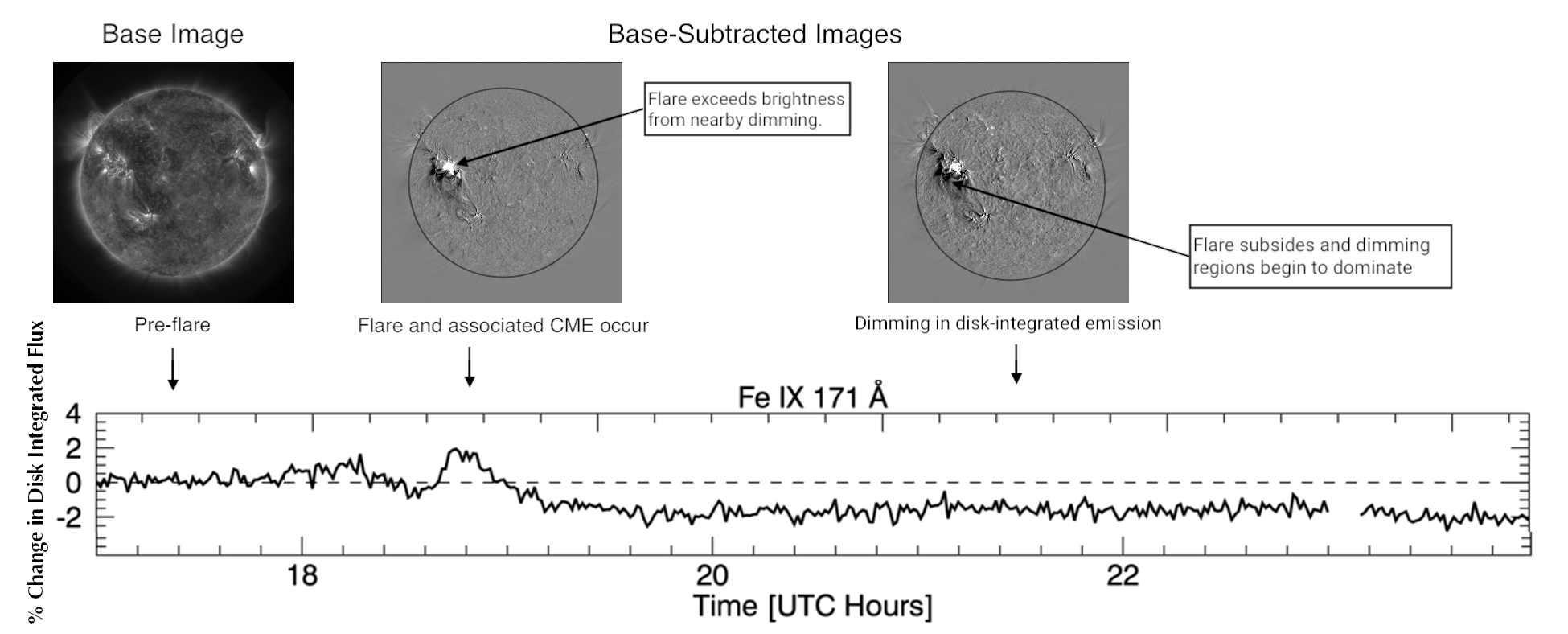}
\caption{Example of coronal dimming from a solar CME.
The upper panels show images from the Solar Dynamics Observatory (SDO; \citealt{pesnell12}) Atmospheric Imaging Assembly (AIA; \citealt{lemen12}) in the 171~\AA\ band, dominated by coronal emission from \ion{Fe}{9}.
The second two panels in the series are differenced from the first to highlight changes in flux.
Below the images is a disk-integrated (``Sun-as-a-star'') lightcurve from the SDO Extreme ultraviolet Variability Experiment (EVE; \citealt{woods12}), also of emission from the \ion{Fe}{9} line at 171~\AA, from \cite{mason14}.
In this example, a flare-associated CME produces transient coronal dimmings that show up in the disk-integrated lightcurve as a drop in coronal emission, indicating that a CME occurred.}
\label{fig:solardimming}
\end{figure*}

\subsection{Coronal Dimming}

A CME is a volume of coronal plasma carrying its own entrained magnetic field that has been expelled from a star.
This phenomenon occurs nearly daily on the Sun, even during solar minimum \citep{yashiro04}.
CMEs often accompany solar flares, but both can occur in isolation.
All solar flares of \textit{GOES} class $>$~X3.1 ($>$~$3.1\sn{-4}$~W~m$^{-2}$ peak 1--8~\AA\ flux at 1 AU) have been accompanied by a CME \citep{yashiro05,harra16}.

A direct, measurable outcome of solar CMEs is coronal dimming.
Ejected coronal material expands as it leaves the sun, rapidly quenching its emission and leaving behind evacuated regions that appear dark against their surroundings \citep{hudson96,shiota05}.
These are likely responsible for features identified as ``coronal depletions,'' ``transient coronal holes,'' and ``coronal dimmings'' throughout the history of observations of the solar corona \citep{hansen74,rust83,hudson96}.
Coronal dimmings were observed for some time as darkened areas in spatially-resolved coronal observations before \cite{woods11} recognized that they produced a measurable decrease in disk-integrated emission as well (i.e., solar irradiance or Sun-as-a-star observations).
Figure \ref{fig:solardimming} illustrates this concept with spatially-resolved and disk-integrated solar data from an example event on the Sun.
Dimming is most pronounced in isolated emission lines at approximately ambient coronal temperatures \citep{woods11,mason14}, with a gradual return to a new quasi-quiescent state over the course of 3-12 hours \citep{reinard08}.

Coronal dimming is diagnostic of solar CMEs, suggesting it could be applied to stars.
In a comparison of disk-integrated observables, \cite{harra16} found that only coronal dimming was a reliable indicator that a CME accompanied a stellar flare.
Other observables tested, including flare duration, active region size, size of flare ribbons, thermal/non-thermal radiation ratios, and properties of chromospheric evaporation were not predictive of CMEs.

\cite{veronig21} applied the coronal dimming method to stars for the first time by analyzing a large set of archival EUV and X-ray observations.\footnote{Observations were from the Extreme Ultraviolet Explorer, XMM-Newton, and Chandra X-ray Observatory.}
They found 21 candidate CME events where coronal emission in at least one band dimmed by $>2\sigma$ following a flare, with depths ranging from 5-56\% and durations of 1-10~h.
This maximum depth is larger than the largest dimmings that have been observed on the Sun \citep[$\approx 10$\%;][]{mason19}, consistent with the expectation that more active stars could produce more massive CMEs.
The range of durations was similar between the Sun and stars, though stellar dimmings skewed toward shorter durations.

In this paper, we present our work to further develop the coronal dimming technique for use with stellar observations.
We provide analytical tools to estimate event masses and accelerations/velocities, highlighting a key strength of the dimming technique: the ability to place limits on the mass of a flare-associated CME in the absence of a dimming detection.
These tools can be used with any disk-integrated, emission-line-resolved data (Sun or star).
Further, we show how coronal dimming can be applied to spectroscopic observations of stellar far ultraviolet (FUV) emission, with results from  an analysis we conducted of archival data of the active K2 dwarf $\mathrm{\epsilon}$ Eridani (\ee).

\section{Analytical Framework for Estimating CME Mass and Velocity}
\label{sec:method}

This section explains the rationale behind simple analytical formulae for making order of magnitude estimates of CME properties from dimming observations.
The derivations below generalize those of \cite{mason16} for estimating masses of solar CMEs from disk-integrated (irradiance) dimming, yielding expressions that can be used for any arbitrary star.
Our derivation deviates in a key point, predicting that $m \propto \mdmax$ in contrast with the $\mMcme \propto \mdmax^{1/2}$ prediction of \citep{mason16}.
Similar derivations have also been carried out for use in estimating masses of solar CMEs from spatially resolved coronal dimming, where the volume of the dimming can be directly constrained from imagery \citep{harrison00,harrison03,zhukov04,aschwanden09}.

\subsection{Mass}
\label{sec:method:mass}
When a body of emitting plasma is ejected from the star, the fractional decline in the emission from the star (the maximum ``dimming depth''), \dmax, is
\begin{equation}
    \mdmax = \frac{F_\mathrm{CME}}{\mFpre},
    \label{eq:d}
\end{equation}
where $F_\mathrm{CME}$ is the flux from the plasma evacuated by the CME and \Fpre\ is the disk-integrated flux of the star before any ejection occurred.
Assuming the coronal material is optically thin at the wavelength of observation, its flux is
\begin{equation}
    F = \int_V{G(T,n_e) n_e n_H dV},
    \label{eq:emint}
\end{equation}
where $V$ represents the volume of emitting material, $n_e$ and $n_H$ are the number densities of electrons and hydrogen, and $G(T,n_e)$ is a (known) function expressing the emissivity of the plasma, which varies with temperature and density.
Assuming a constant temperature and a constant density throughout the emitting volume and setting $n_e \approx n_H \equiv n$ simplifies this to
\begin{equation}
    F = G(T,n) n^2 V.
    \label{eq:em}
\end{equation}

The expression $n^2 V$ is related to the mass of the emitting plasma, $m$, since $m \approx \mu n V$, where $\mu$ is the mean mass of the ions.
Using this to recast the numerator of Eqn. \ref{eq:d} in terms of the emitting mass yields
\begin{equation}
    \mdmax = \frac{n G(T,n) m}{\mu \mFpre},
    \label{eq:d2}
\end{equation}

\begin{figure}
\includegraphics{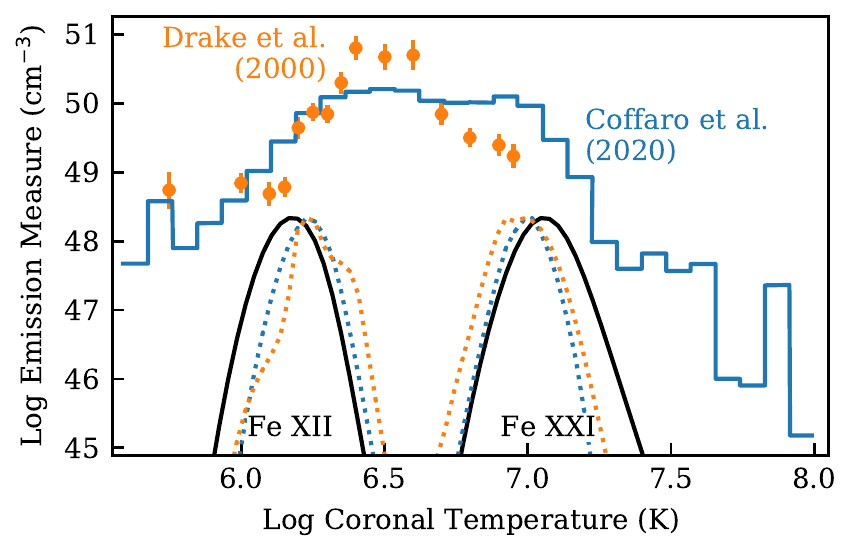}
\caption{Two contrasting emission measure distributions (EMDs) for \ee\ (blue curve and orange dots) compared against the emissivity, $G(T,n)$ (black curves), of \Fexii\ and \Fexxi\ in a solar-metallicity plasma.
The EMD from \cite{coffaro20} is based on X-ray data obtained simultaneous with the FUV data in February 2015.
The emissivities are normalized to peak at the same level and only the top two orders of magnitude are plotted.
Dotted curves show the contribution functions for each ion under the two different EMDs.
Regardless of which EMD is used, the contribution functions peak very near the peak emissivity, justifying the assumption that $G = G_\mathrm{max}$ in our order of magnitude estimates.
}
\label{fig:emd}
\end{figure}

Every component of the above equation except $m$ can be either measured or estimated.
The pre-CME flux and dimming depth (or limit on depth) are direct observables.
Density can be estimated using a density-sensitive line ratio, ideally one that includes the dimming line, but alternatively using a ratio of lines likely to be formed cospatially as indicated by their peak formation temperatures.
The mean atomic weight can be taken to be that of the Sun, as variations in $\mu$ due to stellar metallicity differences contribute negligibly to the overall error budget.
Emissivity can be retrieved from a database such as CHIANTI \citep{dere97}.
The sharp peak in emissivity functions (Figure \ref{fig:emd}) means that, barring extreme slopes of the stellar \edit1{emission measure distribution (EMD)}, the source of the line emission will be dominated by plasma with a temperature near the line's peak formation temperature.
This is the case for \ee, where two disparate reconstructions of its EMD yield contribution functions for the \Fexxi\ and \Fexii\ lines that peak very near the peak of their emissivity functions (Figure \ref{fig:emd}).

That each line only traces plasma within a narrow range of temperatures is a critical point.
A mass estimate from line-dimming will \emph{only} include plasma near the line's peak formation temperature.
The estimate is blind to the ejection of plasma that was too cold or too hot to produce significant emission in the analyzed line.
To make this limitation clear, we refer to the mass in Eqn. \ref{eq:d2} as the ``ejected emitting mass'' (EE mass).
We discuss the implications of this limitation further in Section~\ref{sec:caveats}.

Solving Eqn. \ref{eq:d2} for mass gives
\textbf{
\begin{equation}
    m = \frac{\mu \mdmax \mFpre }{nG(T,n)}.
    \label{eq:m}
\end{equation}}
The inverse relationship with density might seem counterintuitive.
A resolution comes from noting that varying only $n$ means the dimming depth and hence the flux from the would-be CME, $F_\mathrm{CME}$, are held fixed.
Maintaining a constant $F_\mathrm{CME}$ while increasing $n$ necessarily requires decreasing the mass, since the collisional excitations that produce radiation occur more frequently as $n$ increases (up to the point that collisions are so rapid that they de-excite the ions before a radiative transition can occur).

Some readers might find it useful to recast the mass equation in terms of emission measure, $EM = n^2 V$, giving
\begin{equation}
    m = \frac{\mu \mdmax EM_\mathrm{pre}}{n}.
    \label{eq:mem}
\end{equation}
However, we prefer Eqn. \ref{eq:m} because it makes the dependency on the assumed value of $G(T,n)$ explicit.

Whereas our derivation yields $m \propto \mdmax$, that of \cite{mason16} yields $m \propto \mdmax^{1/2}$.
The \cite{mason16} derivation was meant only for use with the Sun and assumes a fixed volume in which dimming occurs.
Variations in dimming depth stem primarily from how effectively plasma is removed from that volume.
Our assumptions are nearly the reverse and intended to be more general for use with other stars.
We assume the dimming volume can vary, but the CME evacuates all plasma from that volume.
Under this framework, variations in dimming depth stem primarily from the total volume that the CME evacuates.

\edit1{The assumption that the CME plasma is fully evacuated does not strongly affect the estimated CME mass.
Introducing an efficiency factor, $f$, into the above formalism, where $f=1$ indicates full evacuation of the plasma, yields
\begin{equation}
    m = \frac{1}{2-f}\frac{\mu \mdmax \mFpre }{nG(T,n)}.
    \label{eq:mmod}
\end{equation}
This lack of sensitivity to $f$ results from a degeneracy between the volume of the evacuated plasma and the drop in density that, together, produce the observed dimming depth.
If plasma is evacuated less efficiently (lower $f$), it must be evacuated from a larger volume to explain the observed drop in emission.
Because the evacuated mass is the product of the drop in density and the volume, these factors nearly cancel.
For a fixed dimming depth, as $f \rightarrow 0$, the evacuated volume blows up, $V \rightarrow \infty$, and $m$ asymptotes to 1/2 of its maximum value.}

For the Sun, the question of which scaling is more accurate can be addressed by empirically fitting measurements of dimming depth for a population of CMEs with estimates of their masses made independent of dimming.
Doing so for the dataset from \cite{mason16}, we found the empirical relationship to have an exponent of $0.84\pm0.09$, falling between the exponents derived here and by \cite{mason16}.
\edit1{In MHD simulations of 7 CMEs and corresponding dimmings, \cite{jin22} recovered a relationship with an exponent of $1.35\pm0.25$ between dimming depth and CME mass.}
A larger dataset will help to resolve whether either or neither scaling is correct.
This could be accomplished by future work using the automatically-extracted events in \cite{mason19}, vetted to \edit1{include} only bona-fide dimmings and cross-matched with the SOHO/LASCO CME catalog \citep{gopalswamy09}.

\subsection{Acceleration, Velocity, and Dimming vs. Time}
Dimming from a CME is not instantaneous.
As the material being ejected moves away from the stellar surface into regions of lower plasma and magnetic pressure, it expands.
The drop in density and any associated adiabatic cooling from the expansion reduce the rate of emission.
This ties the rate of expansion to the rate of dimming, and it means that any functional form for dimming versus time will, implicitly or explicitly, make some assumptions about how the CME moves and expands with time.
\edit1{A CME does not expand into a vacuum, but rather into the ambient environment of the stellar wind.
This affects its movement and expansion, hence the dimming signature.
The global fields and wind properties of active stars differ from the sun \citep[e.g.,][]{jeffers17,wood21} and will affect CME dynamics \citep{alvarado18,alvarado19,alvarado20}.}

As an order-of-magnitude approach, we assume a self-similar CME expansion into circumstellar space, akin to \edit1{\cite{howard82}}.
The CME grows in volume as $V \propto (r/R_\star)^3$, where $r$ is distance from the center of the star to the CME front.
\edit1{We treat the CME as a self-contained plasma such that} the mass, $\mMcme = \mu n V$ is constant, and the density drops as $n \propto (r/R_\star)^{-3}$ as the CME expands outward.
True solar CMEs are comprised of material in, around, and above the erupting filament,
From Eqn. \ref{eq:em}, \edit1{the constant-mass assumption} implies that the emission from the CME drops as
\begin{equation}
    F_\mathrm{CME} = F_\mathrm{CME,0} (r/R_\star)^{-3},
\end{equation}
and the corresponding dimming depth evolves as
\begin{equation}
    \delta(t) = \mdmax (1 - (r/R_\star)^{-3}).
\end{equation}

The $r^{-3}$ dependence means that most of the dimming occurs early in the CME's journey from the star.
The dimming will have reached half maximum by the time the CME has expanded to a height of only $\approx0.25$~\Rstar.
As a result, the early dynamics of the CME are responsible for most of the dimming, with the decline in flux likely tracing the CME's initial acceleration.
Although we expect the CME acceleration to vary in both height and time, as an order-of-magnitude approach we assume the CME undergoes a constant acceleration.
This mimics the fits to CME positions in the SOHO/LASCO catalog of solar CMEs \citep{gopalswamy09}, though those fits are to data above 2~\Rsun.

Assuming a constant acceleration implies that $r(t)=R_\star + at^2/2$ and
\begin{equation}
    \delta(t) = \mdmax \left(1 - \left(1 + at^2/2R_\star\right)^{-3}\right).
    \label{eq:a}
\end{equation}
This enables a fit to a dimming lightcurve, where the measured flux evolves as
\begin{equation}
    F(t) = \mFpre(1 - \delta(t))
    \label{eq:fit}
\end{equation}
The fit takes 3 free parameters: the pre-dimming, pre-flare flux, \Fpre; the maximum dimming depth, \dmax; and the CME acceleration, $a$.

In the LASCO Solar CME catalog, the constant-acceleration $r(t)$ fits are used to extrapolate a uniform velocity at 20~\Rsun, $v_{20}$.
We suggest this same convention be used to assign a CME velocity from a constant-acceleration dimming fit, and that it be fixed to 20~\Rsun\ (rather than 20~\Rstar) to ensure consistent comparisons.
Hence,
\begin{equation}
    v_{20} = \sqrt{2a(20R_\odot - R_\star)} \approx 6\sqrt{aR_\odot}.
    \label{eq:v}
\end{equation}
For a CME undergoing a constant acceleration starting from near the surface of the Sun, the acceleration required to reach typical CME velocities at 20~\Rsun\ are modest relative to surface gravity.
Roughly speaking, $a = $~3~m~s$^{-2}$ yields $v_{20} = 300$~km~\pers\ and $a = $~30~m~s$^{-2}$ yields $v_{20} = 1000$~km~\pers.

\edit1{It is important to note the limitation of assuming a constant acceleration when inferring $v_{20}$.
The true acceleration of a CME will vary with time as the forces ejecting it subside and it moves out of the stellar gravitational potential.
The expansion of the CME as it leaves the star will be controlled by the pressure of the stellar wind plasma and magnetic field \citep[e.g.,][]{alvarado18}, an environment that varies spatially and temporally.
For active stars, ambient conditions will include stronger global fields and denser stellar winds \citep{wood21,reiners22}.
The reader should treat $v_{20}$ as an indication of the magnitude of the velocity a CME could reach, rather than an accurate prediction of the velocity a CME will reach.}

A useful metric for visual estimates is the time to half-depth, $t_{1/2}$.
Under Eqn.~\ref{eq:a}, this is
\begin{equation}
    t_{1/2} = \sqrt{R_\star/2a},
\end{equation}
where we have made the approximation $2^{1/3} - 1 \approx 1/4$.
This makes it possible to quickly gauge what range of accelerations could produce a dimming that reaches most of its depth within an observational visit.
A visit lasting $\sim$10~h would accommodate CMEs from a solar-radius star with $a\gtrsim0.3$~m~s$^{-2}$, $v_{20}\gtrsim90$~km~\pers, very modest values for a solar CME.

This formalism yields a direct relationship between the CME speed at a given distance and the ratio of the dimming slope to the dimming depth, just as in the constant-speed formulation of \cite{mason16}.
Our formulation gives
\begin{equation}
    \delta'(t_{1/2}) \approx \mdmax \sqrt{a/R_\star} \approx \frac{\mdmax v_{20}}{7 \sqrt{R_\star R_\odot}}.
\end{equation}
A typical value from \cite{mason16} for $\delta'/\mdmax$ is 0.0003.
Our equation predicts a corresponding $v_{20} = 1500$~km~\pers.
This is near the upper range of the observed velocities (estimated by position tracking) of solar CMEs analyzed in that work and about a factor of 2 above the median.
Although satisfactory for an order-of-magnitude approach, we believe this offset is an indication that CME accelerations typically decline as the CME leaves the star.

Our derivation deviates in some important respects from the related derivations of \cite{aschwanden09} and \cite{mason16}.
Whereas \cite{aschwanden09} assume a functional form for the time-dependent dimming of a spatially resolved region in the solar coronae (a semi-gaussian), we derive a functional form for the disk integrated dimming based on the assumptions of constant-acceleration and self-similar expansion.
Eqn.~\ref{eq:a} produces a functional form that resembles a logistic function as well as the semi-Gaussian dimming fits prescribed by \cite{aschwanden09}.
\cite{mason16} assume a constant velocity rather than a constant acceleration.
Although little is known about the early acceleration of CMEs \citep{mason16}, the critical importance of the early dynamics of the CME motivate us to choose a constant acceleration rather than an instantaneous acceleration to a constant velocity.
Both formulations yield a dependence of the CME speed on the ratio of dimming slope to depth.
Assuming a constant speed CME results in more rapid dimming, requiring unrealistically low velocities to produce the dimming slopes observed by \cite{mason16}.
This is not an issue for the \cite{mason16} work because the velocity -- dimming slope relationship is empirically calibrated.

The takeaway result of this section is that one can use Eqns.~\ref{eq:a} and \ref{eq:fit} to fit a dimming lightcurve.
This yields a dimming depth and CME acceleration.
The acceleration can be used to estimate the CME speed at a given distance.
The dimming depth, in conjunction with Eqn~\ref{eq:m} and estimates of plasma density (from an appropriate line ratio) and emission measure, yields an order-of-magnitude estimate of the mass of the emitting material that was ejected.

\section{\ee}
\label{sec:star}

\ee\ is a K2 dwarf with an age of 200-800 Myr \citep{janson08} known to host a debris disk \citep{greaves98, macgregor15} and a gas giant planet \citep{hatzes00,mawet19}.
It is often labeled a solar-type star in the literature \citep[e.g., ][]{o22}.
Table \ref{tbl:ee} provides basic properties of the star.

The EMD of \ee's corona has been reconstructed from X-ray and extreme ultraviolet (EUV) observations at a number of epochs \citep{drake00,sanz03,sanz04,coffaro20}.
\ee's EMD peaks in the range of $10^{6.4}-10^{6.8}$~K versus the Sun's peak at $10^{5.8}-10^{6.2}$~K, and at times exhibits a plateau out to $10^7$~K.
Above $10^{6.2}$~K, the emission measure of \ee's corona consistently exceeds the Sun's.
Figure \ref{fig:emd} shows two disparate EMD reconstructions for \ee\ (\citealt{drake00,coffaro20}).
The EMD by \cite{coffaro20} is based on a February~2015 XMM-Newton X-ray observation of \ee\ that occurred simultaneously with the HST data we analyzed.

\cite{coffaro20} found that \ee\ displays an X-ray activity cycle in sync with its known $2.92\pm0.02$~yr chromospheric activity cycle.
In chromospheric emission, there is evidence for a $\approx13$~yr cycle as well \citep{metcalfe13,jeffers22}.
The observations described in Section~\ref{sec:data} coincide with minima in the activity cycle of \ee.

%%%%%%%%%%%%%%%%%% TABLE EE Props %%%%%%%%%%%%%%%%%%%%%%%
\begin{deluxetable}{lcc}
\caption{Properties of \ee.}
\tablewidth{\textwidth}
\tablehead{\colhead{Parameter} & \colhead{Value} & \colhead{Reference}}
\startdata
Mass & $0.81\pm0.02$ \Msun & \cite{rosenthal21}\\
Radius & $0.76\pm0.01$ \Rsun & \cite{rosenthal21}\\
\Teff\ & $5020\pm58$ K & \cite{rosenthal21}\\\relax
[Fe/H] & $-0.04\pm0.06$ & \cite{rosenthal21}\\
Distance & $3.203\pm0.005$ pc & GC18\tablenotemark{b}\\
Gaia G Mag & $3.477\pm0.006$ & GC16\tablenotemark{b}\\
$\log(T)$ at & 6.5 $\log(\mathrm{K})$ & \cite{coffaro20} \\
  EMD Peak & & \\
Age & 200-800 Myr & \cite{janson08}\\
\Prot\tablenotemark{a} & $11.68\pm0.6$ d & \cite{donahue96}\\
\enddata
\tablenotetext{a}{Period agrees with photometry from the Transiting Exoplanet Survey Satellite \citep{ricker15}.}
\tablenotetext{b}{\cite{gaia16,gaia18}}
\label{tbl:ee}
\end{deluxetable}
%%%%%%%%%%%%%%%%%%%%%%%%%%%%%%%%%%%%%%%%%%%%%%%%%%%%%%%%%

\begin{figure*}
\includegraphics{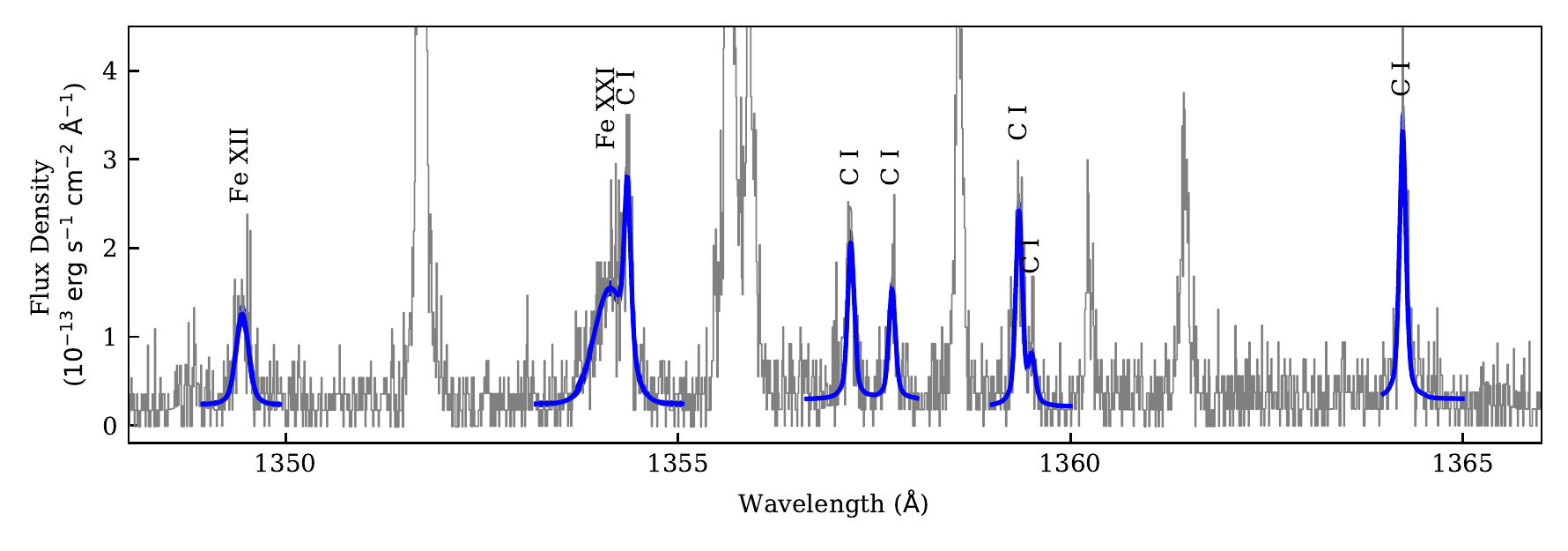}
\caption{Simultaneous fit to \Ci\ and \Fexxi\ lines for deblending.
We constrained the blended \Ci\ profile to have the same broadening as the unblended \Ci\ lines.
The thickness of the model curves (blue) represents uncertainty.}
\label{fig:c1fit}
\end{figure*}

\section{\ee\ Observations \& Reduction}
\label{sec:data}

We retrieved data from three separate HST visits of \ee\ that took place in February 2015, March 2017, and August 2017.
The February 2015 visit was part of the MUSCLES Treasury Survey, program 13650 \citep[PI Kevin France]{france16}.
The 2017 visits, programs 14909 (PI Thomas Ake) and 15365 (PI Rachel Plesha) were conducted for instrument calibration, specifically to obtain dispersion solutions using the high SNR emission lines from \ee.
We verified that the reduced data from these visits have an accurate dispersion solution (line centers match during quiescence to within the stated 7.5~\kms\ wavelength accuracy of COS; \citealt{plesha19}).
We did not incorporate exposures that used the 1222 and 1223 central wavelength (CENWAVE) settings during the August 2017 visit into the search for dimming signals.
These were brief observations with rarely used modes and we found indications that the flux calibration of the 1222 data were inaccurate.
However, we did use the 1223 observation to constrain the density of \Fexii\ emitting plasma as described in Section~\ref{sec:analysis}.
Aside from these exposures, the data analyzed include everything obtained with the G130M grating prior to 2021.
The data are available at MAST with DOI \dataset[10.17909/3sy1-zm28]{\doi{10.17909/3sy1-zm28}}.
Table \ref{tbl:obs} provides a summary of the analyzed observations.
% DOI info copied from MAST:
%  DOI: 10.17909/3sy1-zm28
%   Dataset Title: Eps Eri Coronal Dimming of Fe Lines from CMEs

\begin{deluxetable}{ccccc}
\caption{Analyzed Observations of \ee.}
\tablewidth{\textwidth}
\tablehead{\colhead{PI} & \colhead{Date} & \colhead{Start} & \colhead{Duration} & \colhead{CENWAVE\tablenotemark{a}}\\
 & & \colhead{UT} & \colhead{s} & }
\startdata
France & 2015-02-02 & 08:18:30 & 2275 & 1291\\
 &  & 09:40:39 & 2712 & 1291\\
 &  & 11:16:11 & 2712 & 1318\\
 &  & 12:54:58 & 2712 & 1318\\
 &  & 14:32:00 & 2712 & 1318\\
Ake & 2017-03-07 & 06:33:08 & 928 & 1300\\
 &  & 06:51:47 & 929 & 1318\\
 &  & 07:57:19 & 2722 & 1291\\
 &  & 09:33:10 & 2722 & 1309\\
 &  & 11:12:00 & 2722 & 1327\\
Plesha & 2017-08-11 & 11:59:19 & 2073 & 1300\\
 &  & 13:20:32 & 2748 & 1291\\
 &  & 14:55:54 & 2748 & 1309\\
 &  & 16:31:16 & 2748 & 1327\\
 & 2017-08-12 & 11:50:03 & 2067 & 1318\\
 &  & 14:46:28 & 2748 & 1223
\enddata
\tablenotetext{a}{Central wavelength setting (Å, approximate).}
\tablecomments{All observations are from the Cosmic Origins Spectrograph with the G130M grating. }
\label{tbl:obs}
\end{deluxetable}

We began the data analysis by constructing lightcurves of the emission from the bright, flare-sensitive emission lines of \Cii\ (1335~\AA) and \Siiv~(1393,~1402~\AA) following the methodology of \cite{loyd18}.
This revealed a single prominent flare in each visit, along with a smattering of smaller flares.
Each flare is generated by magnetic reconnection that could also have produced a CME, potentially causing the coronal emission lines of \Fexii\ and \Fexxi\ to dim.

\begin{figure}
\includegraphics{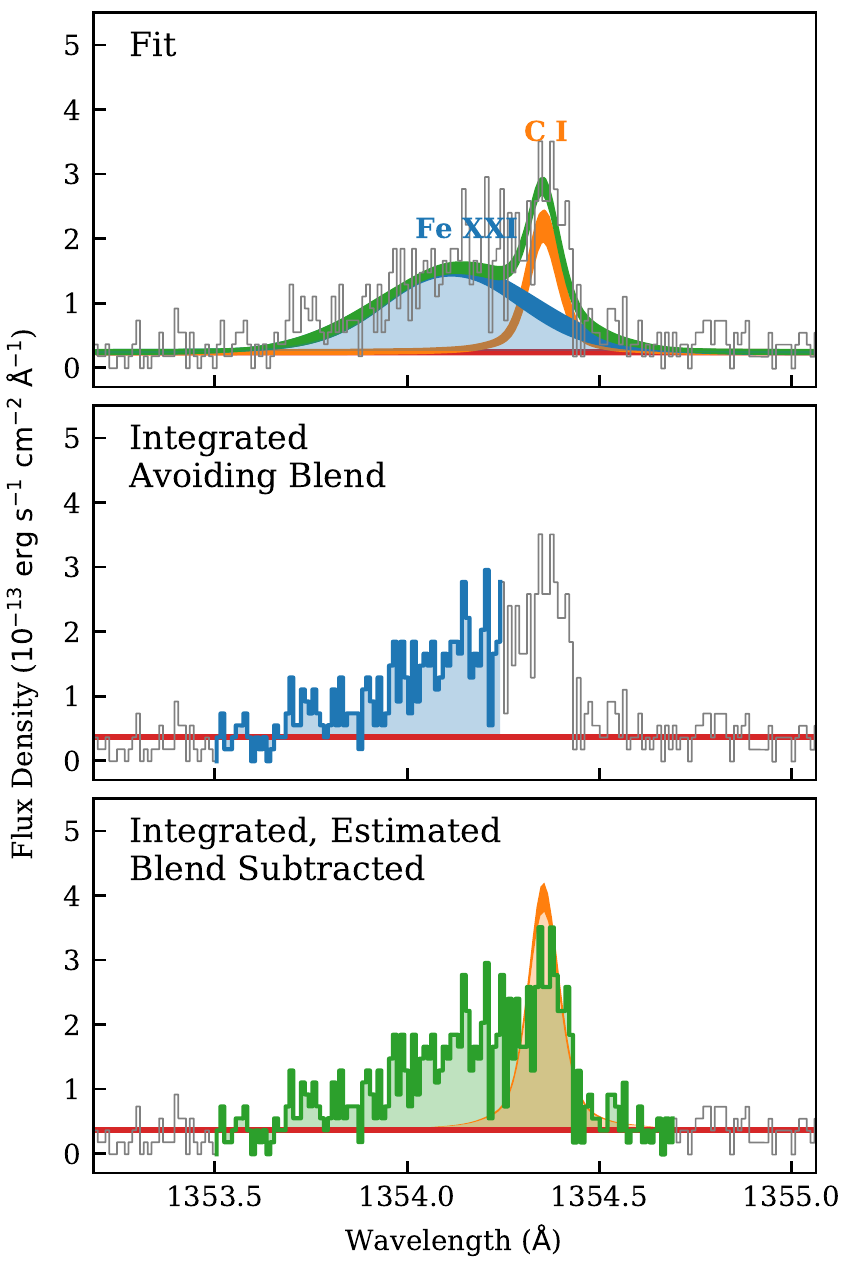}
\caption{Demonstration of the three methods used to integrate the \Fexxi\ line.
The profiles here correspond to the fifth set of points in Figure \ref{fig:lc}.
The \Ci\ line shown in the bottom panel is meant only to represent the subtracted flux, no line fit was actually made.
Curve thickness represents uncertainty in the top panel.}
\label{fig:fexxi}
\end{figure}

\begin{figure}
\includegraphics{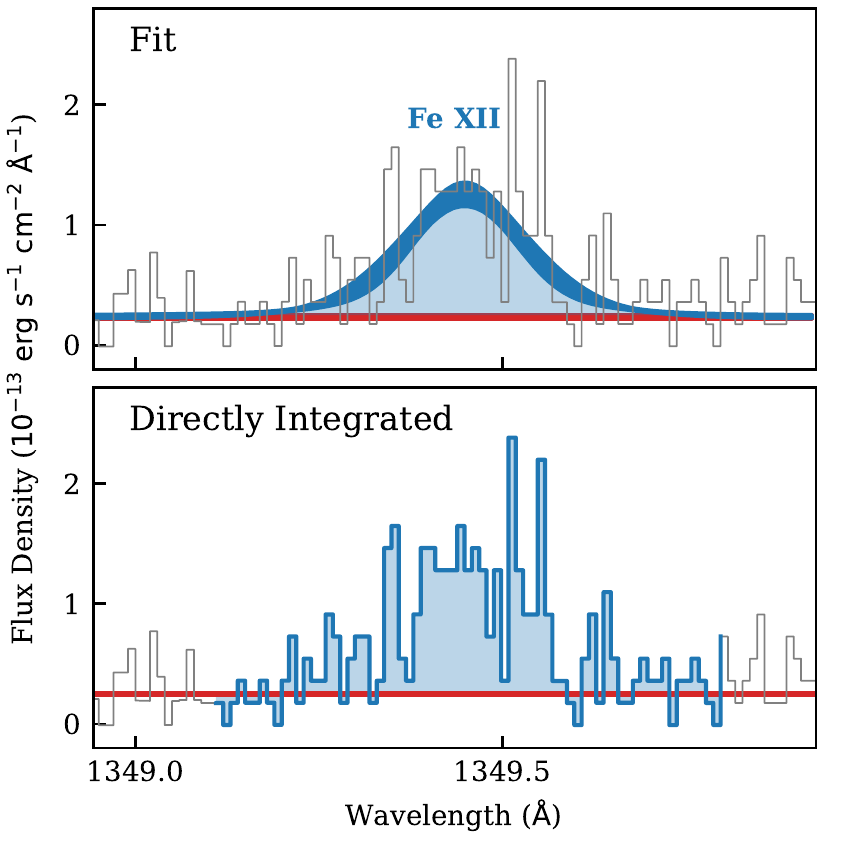}
\caption{Demonstration of the two methods used to integrate \Fexii.
The profiles here correspond to the fifth set of points in Figure \ref{fig:lc}.
Curve thickness represents uncertainty in the top panel.}
\label{fig:fexii}
\end{figure}

\Fexii\ (1242 and 1349~\AA) and \Fexxi\ (1354~\AA) are weak iron lines resulting from forbidden magnetic dipole transitions.
\cite{ayres03} identified these lines in the FUV spectra of several active stars.\footnote{We omit the bracket notation for forbidden lines (e.g., [\Fexii]) throughout this paper because we found it visually distracting.}
Their peak formation temperatures are 1.6~MK (\Fexii) and 11~MK (\Fexxi) \citep{landi12}.
In the spectrum of \ee, their integrated flux is 10-50$\times$ less than that of bright emission lines like the aforementioned \Siiv.
\Fexii~1242~\AA\ could not be analyzed for dimming because the portion of the detector capturing that line was powered off in the exposures we analyzed to avoid count rate limit violations from the star's bright \lya\ line.
Table~\ref{tbl:lines} provides the properties of all the emission lines utilized in this work.
% import spectralPhoton as sp
% spec = sp.Spectrum.read_muscles(rc.path_coadd)
% spec.plot()
% spec.integrate((1241.5,1242.3)*u.AA)
% spec.integrate((1392,1394)*u.AA)
% spec.integrate((1335.3,1336.5)*u.AA)

%%%%%%%%%%%%%%%%%%%% TABLE LINES %%%%%%%%%%%%%%%%%%%%%%%%
\begin{deluxetable*}{lccccccccl}
\rotate
\scriptsize
\caption{Emission lines used in this work.}
\tablewidth{\textwidth}

\tablehead{\colhead{Ion} & \colhead{$\lambda_\mathrm{rest}$} & \colhead{$A_ij$} & \colhead{Transition} & \colhead{Flux\tablenotemark{b}} & \colhead{FWHM} & \colhead{$\log(T_\mathrm{peak})$\tablenotemark{c}} & \colhead{$\log(n_\mathrm{quench})$} & $G(T,n)_\mathrm{max}$ & Used For\\
\colhead{} & \colhead{\AA} & \colhead{\pers} & \colhead{Type\tablenotemark{a}} & \colhead{erg \pers\ \percmsq} & \colhead{km \pers} & \colhead{K} & \colhead{\percmcb} & \colhead{$\mathrm{erg\ s^{-1}\ cm^3}$} & }
\startdata
\Fexii & 1242.00 & $3.17\sn{2}$ & M1 & $6.6\sn{-15}$ & 50 & 6.2 & 10.9 & $7.2\sn{-27}$ & Density\\
 & 1349.40 & $1.73\sn{2}$ & M1 & $3.0\sn{-15}$ & 50 & 6.2 & 10.5 & $3.5\sn{-27}$ & Dimming \& Density \\
\Fexxi & 1354.08 & $6.49\sn{3}$ & M1 & $1.5\sn{-15}$\tablenotemark{d} & 130 & 7.1 & 12.6 & $3.9\sn{-27}$ & Dimming\\
\Ci & 1354.28 & $3.0\sn{7}$ & E1 & $4.5\sn{-15}$ & 17 & & & & Deblend 1354 \AA\\
 & 1357.13 & $1.4\sn{7}$ & E1 & $2.6\sn{-15}$ & 17 & & & & Deblend 1354 \AA\\
 & 1357.67 & $1.1\sn{6}$ & E1i & $1.3\sn{-15}$ & 17 & & & & Deblend 1354 \AA\\
 & 1359.28 & $2.2\sn{6}$ & E1i & $2.9\sn{-15}$ & 17 & & & & Deblend 1354 \AA\\
 & 1359.44 & $9.7\sn{5}$ & E1i & $8.4\sn{-16}$ & 17 & & & & Deblend 1354 \AA\\
 & 1364.16 & $1.6\sn{7}$ & E1 & $4.5\sn{-15}$ & 17 & & & & Deblend 1354 \AA\\
\Nv & 1238.82 & $3.40\sn{8}$ & E1 & $1.1\sn{-13}$ & 48 & 5.3 & & & Deblend 1242 \AA\\
 & 1242.80 & $3.37\sn{8}$ & E1 & $6.3\sn{-14}$ & 48 & 5.3 & & & Deblend 1242 \AA\\
\Cii & 1334.53 & $2.41\sn{8}$ & E1 & $1.9\sn{-13}$\tablenotemark{e} & \nodata\tablenotemark{f} & 4.6 & & & Flare Identification\\
 & 1335.71 & $2.88\sn{8}$ & E1 & $4.9\sn{-13}$ & \nodata\tablenotemark{f} & 4.6 & & & Flare Identification\\
\Siiv & 1393.76 & $8.80\sn{8}$ & E1 & $2.4\sn{-13}$ & \nodata\tablenotemark{f} & 4.9 & & & Flare Identification\\
 & 1402.77 & $8.63\sn{8}$ & E1 & $1.3\sn{-13}$ & \nodata\tablenotemark{f} & 4.9 & & & Flare Identification\\
\enddata
\tablenotetext{a}{E1: electric dipole, i: intercombination, M1: magnetic dipole}
\tablenotetext{b}{Computed using coadded spectrum including flares.}
\tablenotetext{c}{Error bars give the range where the emissivity drops an order of magnitude from the peak value, providing a gauge for the range of plasma temperatures the line could plausibly trace.}
\tablenotetext{d}{Out-of-flare average from the 2017 epochs.}
\tablenotetext{\edit1{e}}{\edit1{Affected by absorption from the interstellar medium.}}
\tablenotetext{f}{We did not fit profiles to these lines.}
\label{tbl:lines}
\end{deluxetable*}

Because the Fe lines are weak, we hand-selected sub-exposure intervals over which to measure their flux.
The intervals provide at least a few points (of unequal duration) before and during each flare, with greater sampling after each flare.
These intervals were then fed to the calCOS pipeline (version 3.2.1) using the latest reference files from the Calibration Reference Data System (CRDS; version 10.3.0) as of 2021 September.
This yielded pipeline-extracted spectra for each sub-exposure interval.

\Fexxi~1354~\AA\ is blended with a \Ci\ line \citep{ayres03}, complicating measurements of its flux, especially during flares.
The crux of the data analysis was obtaining accurate measurements of relative changes in the \Fexxi\ line flux at sub-exposure intervals, particularly given that the ratio of the line and its blend will vary.
We employed three separate methods to measure the \Fexxi\ line flux so that we could check for agreement between different data reduction methods and pick the method that yields the least scatter.

Our first method of measuring the \Fexxi\ flux was through simultaneous fitting of the line and blend profiles.
This was the most sophisticated and computationally intensive method.
We included 4 nearby \Ci\ lines in the fits to help converge on the correct line profile of the blended \Ci\ line.
The broadening of the \Ci\ lines was held fixed between components (i.e., it could only vary globally).
Meanwhile, the flux of each component could vary independently.
We used a double Gaussian profile for the \Ci\ lines because this provided a markedly better fit to the wings of these and other strong emission lines in tests on full exposures.
The Fe lines were adequately fit by a single Gaussian profile.
We broadened all modeled lines with the COS line spread function.
The model also included a constant for continuum flux, allowed to take a different value in each band of wavelengths used in the fits.
Figure~\ref{fig:c1fit} displays a fit to an example exposure over the full range of the bands used.
To estimate uncertainties, we used \texttt{emcee}\footnote{https://emcee.readthedocs.io} \citep{foreman13} to explore the posterior of the fits, sampling to a minimum chain length of 100$\times$ the maximum sample autocorrelation length among the walkers.
The top panel of Figure~\ref{fig:fexxi} zooms in on the \Fexxi\ blend, showing the fit profiles of \Fexxi, \Ci, the continuum constant, and the sum of these components.

Our second method of measuring \Fexxi\ flux was a partial integration of the observed line flux, avoiding the \Ci\ blend.
We used adjacent bands free of lines to estimate and subtract an average continuum value and computed Poissonian errors based on total counts in the signal, background, and continuum regions.
Using Poissonian errors avoids inflated uncertainties that result from propagating the calCOS-supplied uncertainties on flux density that are subject to an elevated uncertainty floor (see, e.g., extended data Figure 4 of \citealt{ben22}).
Poissonian errors are adequate given that the dimming signal is a relative change in flux, not absolute.
The middle panel of Figure~\ref{fig:fexxi} illustrates this method, with the integrated range filled and the continuum value shown as a red line.

Our third method of measuring \Fexxi\ flux was to directly integrate the entire blend, then subtract an estimate of the \Ci\ component's flux.
To estimate the flux of the blended \Ci\ component in each sub-exposure, we multiplied the integrated flux of the nearby, unblended, \Ci\ lines by the ratio of the blended line flux to that of the unblended lines.
The line ratio multiplier came from fitting line profiles to a coadded spectrum combining all exposures.
From that fit, we computed the ratio of the flux in the blended \Ci\ line to the unblended lines.
When integrating the blended \Fexxi\ flux and unblended \Ci\ fluxes in sub-exposures, we subtracted a continuum flux estimated from adjacent line-free ranges and computed Poissonian uncertainties.
To account for potential astrophysical variability in the ratio of line fluxes, we computed the scatter in the ratios of unblended \Ci\ components and incorporated this scatter into the uncertainty of the blend-subtracted \Fexxi\ flux.
The bottom panel of Figure~\ref{fig:fexxi} illustrates this method, representing the subtracted \Ci\ flux as a Gaussian profile, although in reality we made no assumption about the line profile other than in the fit to the coadded data.

The \Fexii~1349~\AA\ line exhibits no significant blends.
Hence, we measured the flux of this line using only two methods: by fitting the line profile or by direct integration.
Figure \ref{fig:fexii} illustrates these two methods.

We used the line flux measurements to generate lightcurves of \Fexxi\ and \Fexii\ emission.
Figure~\ref{fig:lc} shows the resulting lightcurves for the 2015 visit.
The three flux estimates generally agree within uncertainties.
For \Fexxi, we favor the measurements where we estimated the \Ci\ blend assuming a constant line ratio, and, for \Fexii, we favor the measurements made by direct integration.
These methods exhibit lower scatter than the line profile fits.
The resulting lightcurves form the basis for our subsequent dimming analysis.

\section{Analysis}
\label{sec:analysis}

\begin{figure*}
\includegraphics{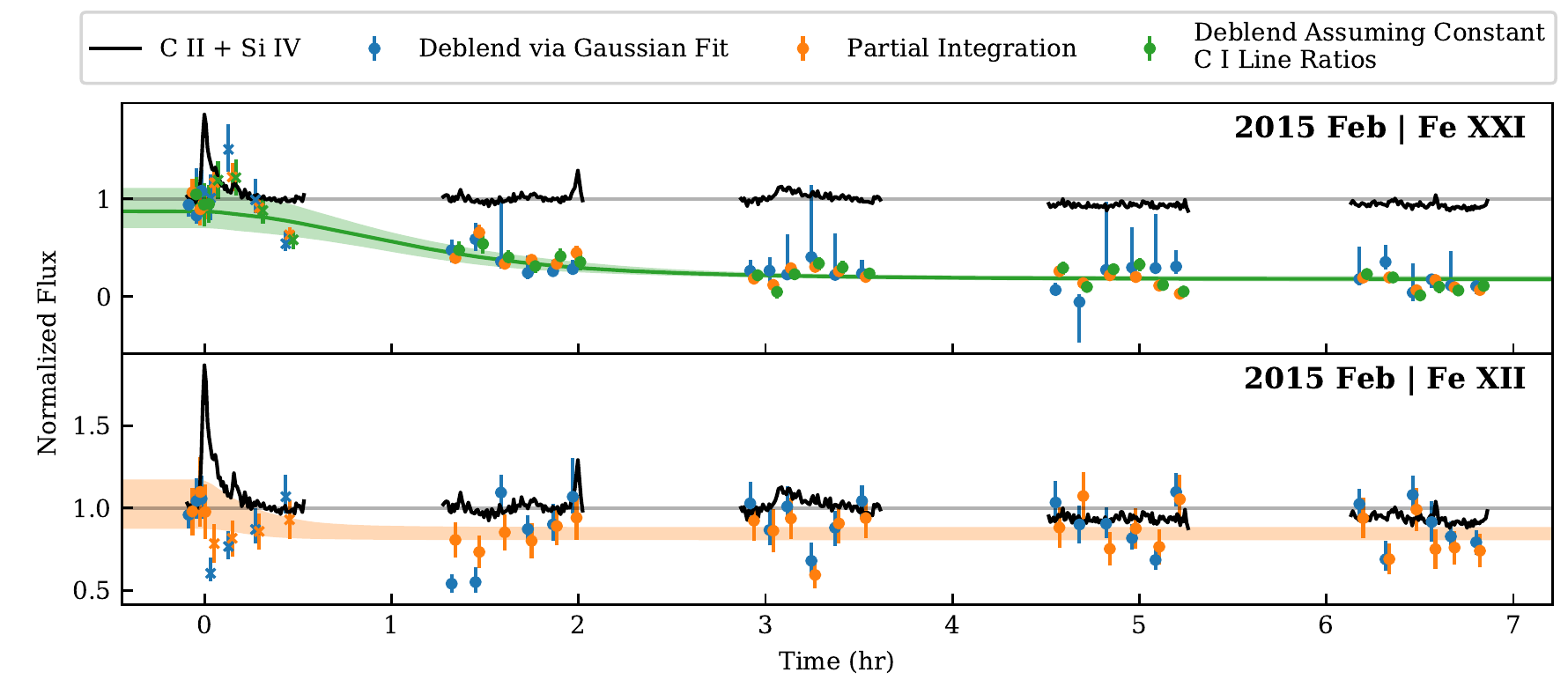}
\caption{February 2015 epoch, lightcurves of the HST observations from program 13650.
Point colors indicate the different methods used to measure the Fe line fluxes as demonstrated in Figures \ref{fig:fexxi} and \ref{fig:fexii}.
The black line\edit1{s} are reference lightcurves from summing the flux of two of the brightest FUV lines.
Possible dimming profiles explored by the MCMC sampler are shown as the translucent regions matched to the color of the points fit, with the best fit for \Fexxi\ shown as the solid green line.}
\label{fig:lc}
\end{figure*}

With \Fexii\ and \Fexxi\ fluxes in hand, we fit the time evolution of those fluxes during each of the three observing visits with the form given in  Eqns.~\ref{eq:a} and \ref{eq:fit}.
We added a Gaussian scatter hyperparameter to accommodate astrophysical white noise in the lightcurves.
The fit to the 2015 visit is shown in Figure~\ref{fig:lc}.
As with the fits to line profiles, we used \texttt{emcee} to explore the posterior of the fit and establish uncertainties or limits on fit parameters.
We sampled fits until the number of steps reached 100$\times$ the maximum sample autocorrelation length among the walkers. %CHECKME

Constraints on the fits are as follows.
We fixed the start time of the dimming to be the start of the \Cii~$+$~\Siiv\ flares in the 2015 and 2017 March visits and the start of the first (smaller) of the two adjacent flares in the 2017 August visit.
This choice prevented the sampler from exploring solutions the data could not reasonably constrain (e.g., a dimming that begins before the first exposure), enabled faster sampling, and facilitated the interpretation of resulting constraints.
Points in the exposure following the flare and any exposure thereafter where the \Cii~$+$~\Siiv\ flux was elevated were masked from the fit.
The acceleration, $a$, was restricted to a uniform prior with a ceiling of $1000$~m~s$^{-2}$ and a floor that ensured the dimming (if present) would reach half depth prior to the end of the observing visit.
In the context of solar events, this was an unrestrictive acceleration prior.
Preventing arithmetic overflow errors motivated us to impose a uniform prior between 0 and 10 on the relative value of the white noise hyperparameter.
We used a uniform prior between 0 and 1 for fractional dimming depth, \dmax.

To derive constraints on the \Fexxi\ and \Fexii\ EE masses and velocities of the CMEs, we applied Eqns.~\ref{eq:m} and \ref{eq:v}, as described in Section~\ref{sec:method}, to the MCMC chains of fit parameters.
We assumed the plasma temperature was that of peak emission for the line considered and that the mean atomic weight was that of the Sun.

Our approach to estimating the plasma density differed between \Fexxi\ and \Fexii.
For \Fexxi, we used a plasma density of $\log(n_e) = 11.02\pm1.15$~\percmcb\ measured by \cite{ness02} from Chandra observations of the \ion{Ne}{9} line, which has a formation temperature of 10~MK, very near the 11~MK formation temperature of \Fexxi.
To account for uncertainty in this value, we sampled random values from a log-normal distribution representing the measured value and its uncertainty.

For \Fexii, we directly constrained the plasma density using the 2017 August observations made with the 1223 CENWAVE setting.
This mode placed the \lya\ line in a detector gap, enabling both the 1242~\AA\ and 1349~\AA\ components of the \Fexii\ line to be measured.
The ratio of these lines is sensitive to plasma density between the values of $10^{9.5} - 10^{11.5}$~\percmcb.
We fit the spectrum from the 1223 observation using Gaussian line profiles for the two \Fexii\ components present and two-component Gaussian profiles for the \Nv\ 1238,1242~\AA\ doublet that is marginally blended with the \Fexii~1242~\AA\ line, following the same methodology as with the \Fexxi\ deblending fits described in Section~\ref{sec:data}.
We then used fluxes sampled from the MCMC fits to compute the density consistent with the flux ratio with the \texttt{CHIANTIpy} package\footnote{https://chiantipy.readthedocs.io} \citep{dere97,zanna21}.
This yielded a density of $\log{n_e} = 10.5\pm0.2$.
We sampled directly from the resulting MC chain of densities when constraining the EE mass associated with possible \Fexii\ dimmings.

\begin{figure}
\includegraphics{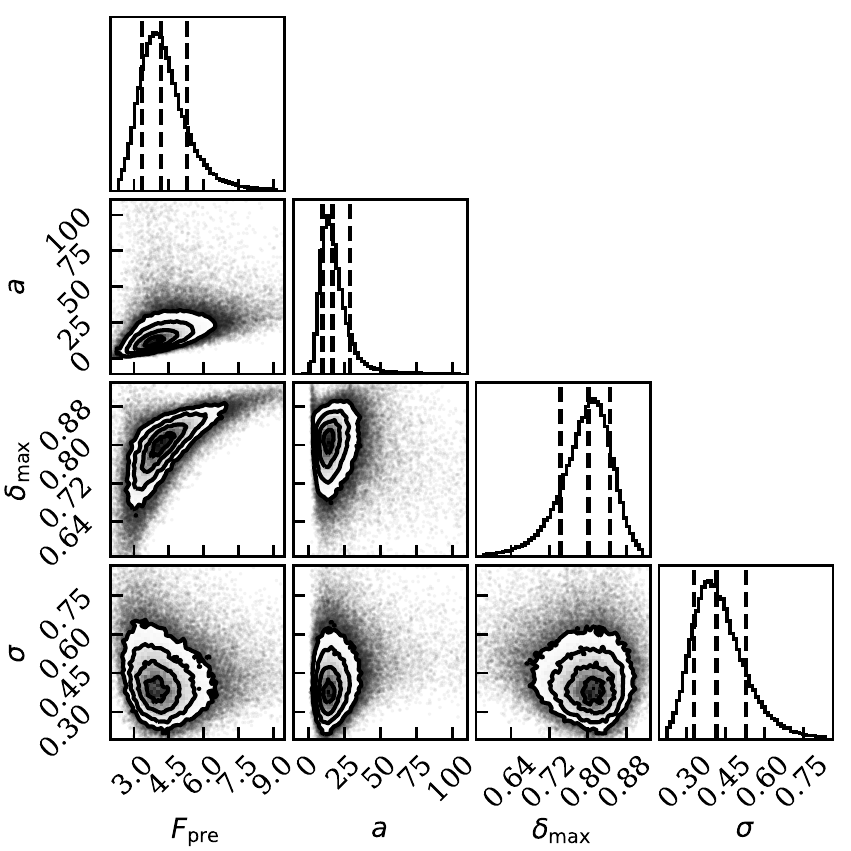}
\caption{February 2015 epoch, MCMC chains from the fit to the \Fexxi\ data in Figure \ref{fig:lc}. \fitunits}
\label{fig:chainxxi}
\end{figure}

\begin{figure}
\includegraphics{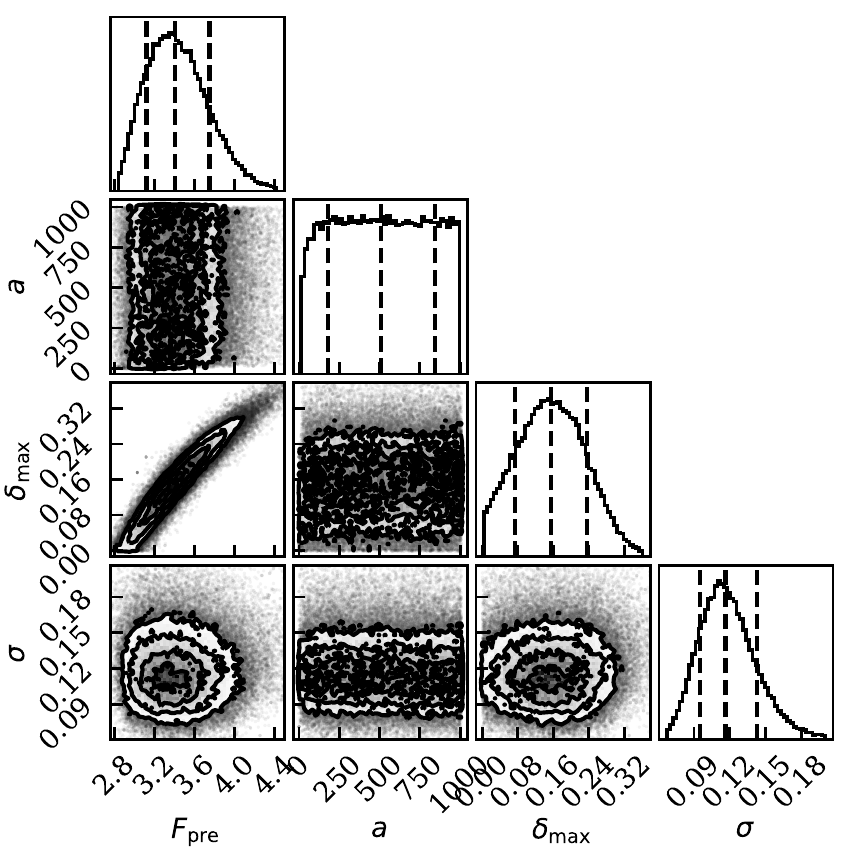}
\caption{February 2015 epoch, MCMC chains from the fit to the \Fexii\ data in Figure \ref{fig:lc}. \fitunits}
\label{fig:chainxii}
\end{figure}

\begin{figure}
\includegraphics{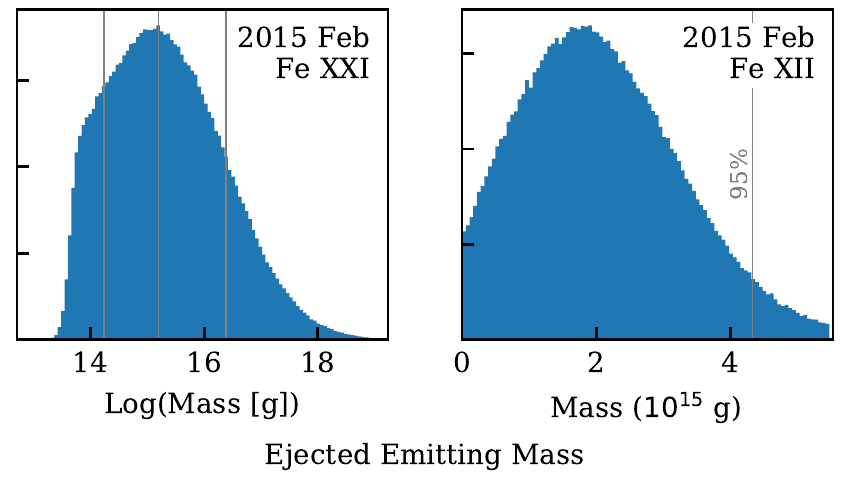}
\caption{Derived ejected emitting masses that could explain the data following the flare in Figure~\ref{fig:lc} if a CME occurred.
Note the disparate axis scales: logarithmic on the left, linear on the right.
\linedesc}
\label{fig:masses}
\end{figure}

\begin{figure}[t]
\includegraphics{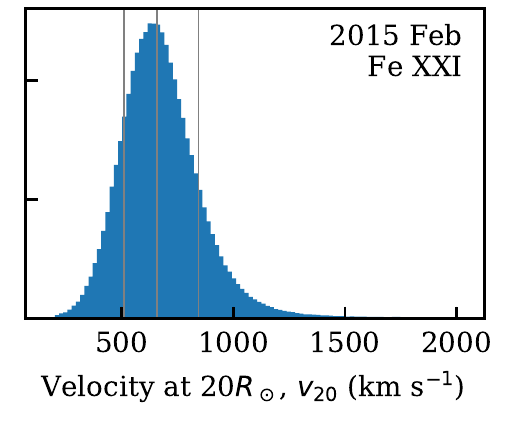}
\caption{CME velocities at 20$R_\odot$ that could explain the drop in \Fexxi\ emission following the flare in Figure~\ref{fig:lc}, assuming constant acceleration and self-similar expansion.
\edit1{Variable acceleration and expansion, influenced by ambient stellar wind conditions and not probed by this technique, will modify the true velocity at 20~\Rsun.}
\linedesc}
\label{fig:velocities}
\end{figure}

\begin{figure*}
\includegraphics{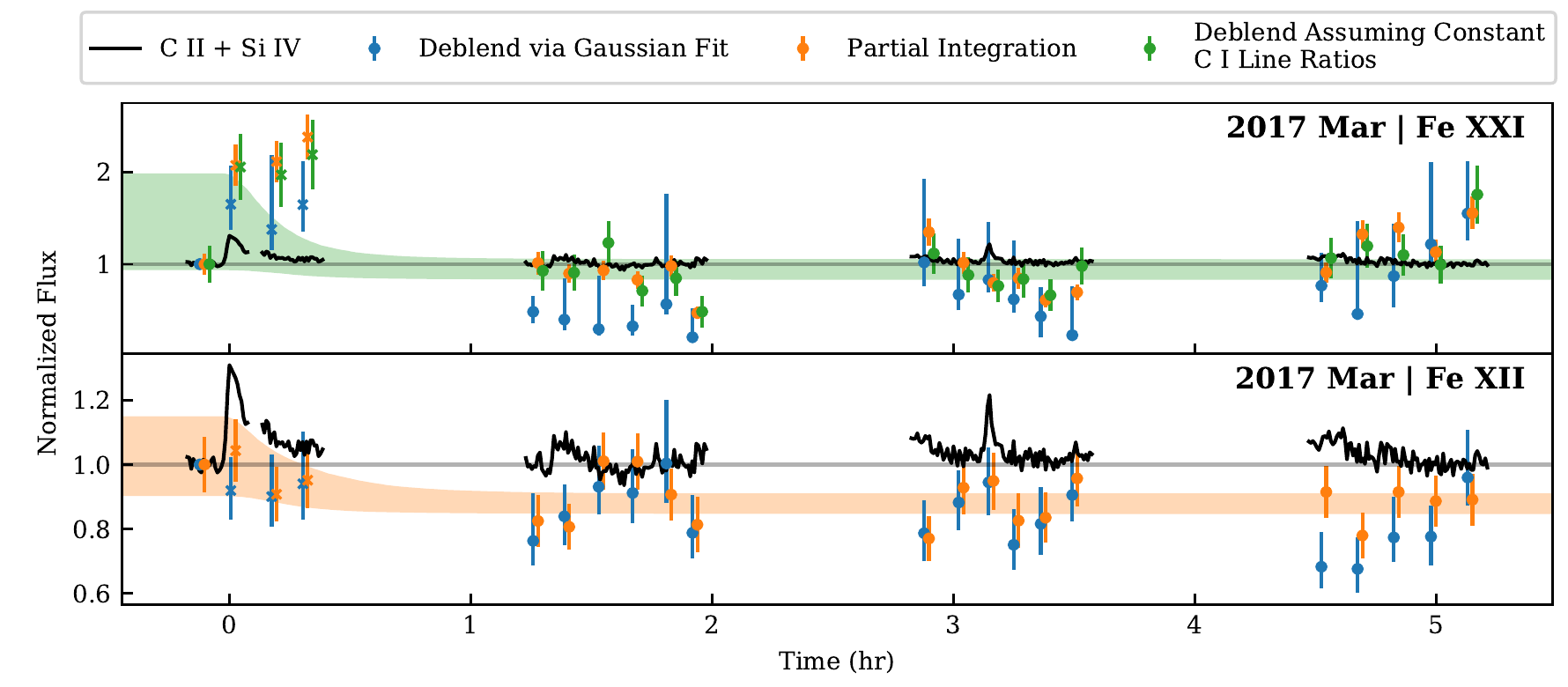}
\caption{March 2017 epoch, lightcurves of the HST observations from calibration program 14909.
Point colors indicate the different methods used to measure the Fe line fluxes as demonstrated in Figures \ref{fig:fexxi} and \ref{fig:fexii}.
The black line are reference lightcurves from summing the flux of two of the brightest FUV lines.
Possible dimming profiles explored by the MCMC sampler are shown as the translucent regions, matched to the color of the points fit.}
\label{fig:blc}
\end{figure*}

\begin{figure}
\includegraphics{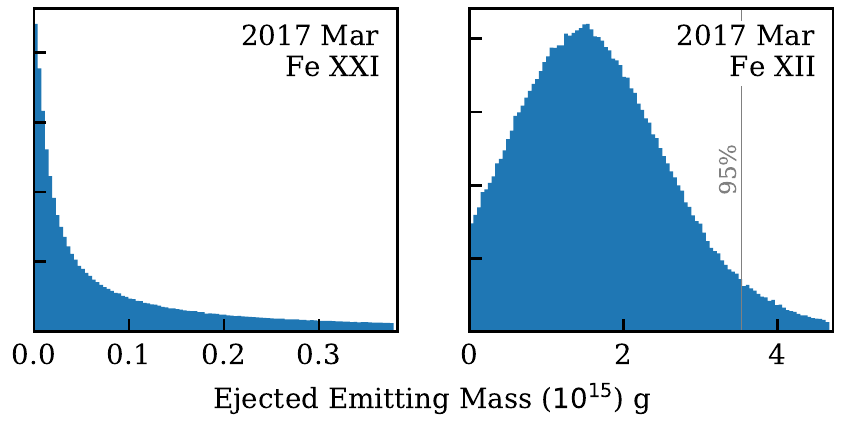}
\caption{Derived ejected emitting masses that could explain the data in Figure \ref{fig:lc} if a CME occurred. The left distribution is heavy-tailed and the 95\% limit is beyond the plot range.
\linedesc}
\label{fig:bmasses}
\end{figure}

\begin{figure*}
\includegraphics{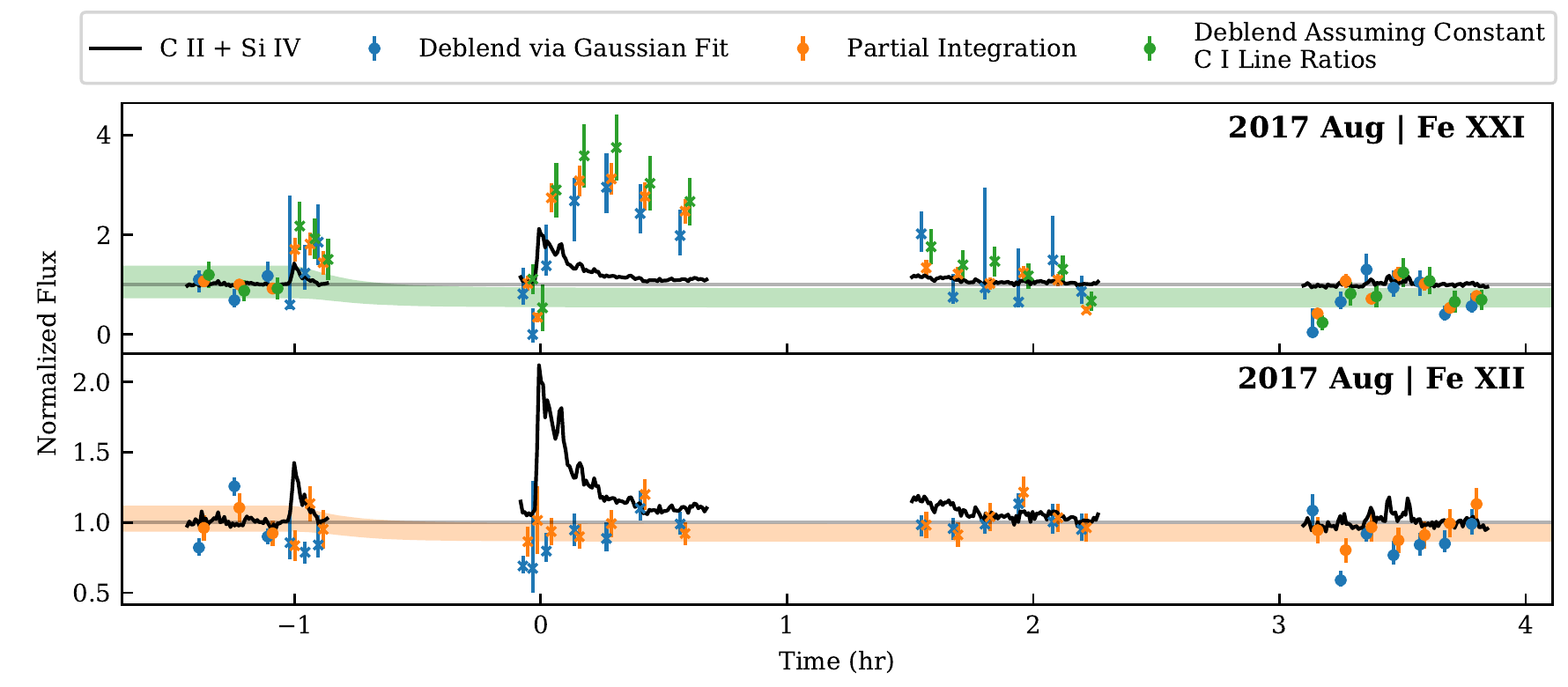}
\caption{August 2017 epoch, lightcurves of the HST observations from calibration program 15365.
Point colors indicate the different methods used to measure the Fe line fluxes as demonstrated in Figures \ref{fig:fexxi} and \ref{fig:fexii}.
The black lines are reference lightcurves from summing the flux of two of the brightest FUV lines.
Possible dimming profiles explored by the MCMC sampler are shown as the translucent regions, matched to the color of the points fit.}
\label{fig:clc}
\end{figure*}

\begin{figure}
\includegraphics{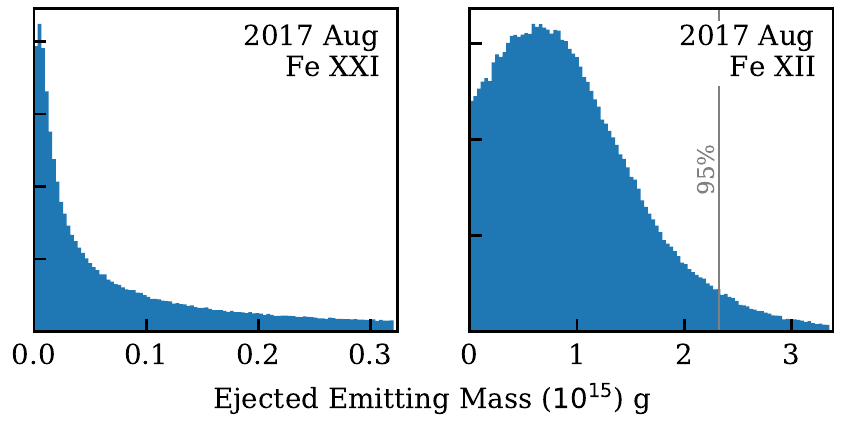}
\caption{Derived ejected emitting masses that could explain the data in Figure \ref{fig:lc} if a CME occurred. The left distribution is heavy-tailed and the 95\% limit is beyond the plot range.
\linedesc}
\label{fig:cmasses}
\end{figure}

\section{Results}
\label{sec:results}
The primary results of this analysis are light curves, fits of the dimming models to those light curves, and the EE mass estimates inferred from those fits.
Since the data include \Fexxi\ and \Fexii\ emission from three separate epochs, each with a dominant flare, we conducted a total of six fits.

We show results from each step of this process for the 2015 epoch.
The lightcurves with model fits overplotted are shown in Figure~\ref{fig:lc}.
Posteriors on the model fit parameters are shown as corner plots in Figure~\ref{fig:chainxxi} (\Fexxi) and Figure~\ref{fig:chainxii} (\Fexii).
The \Fexxi\ emission clearly declines following the flare, with a Bayes factor $\gg30$ relative to no decline ($\delta = 0$).
There is no clear dimming in the \Fexii\ data.
Although a 15\% dimming produces the best fit, the Bayes Factor relative to no dimming is 0.4, favoring the null hypothesis of no dimming.

Figure~\ref{fig:masses} shows the resulting posteriors on EE mass for the 2015 event using the \Fexxi\ and \Fexii\ lightcurves.
The \Fexxi\ posterior on EE mass shows a low-mass edge, the cause of which is discussed in Section~\ref{sec:caveats}.
Figure~\ref{fig:velocities} shows the corresponding posterior on the 20$R_\odot$ velocity of the CME based on the \Fexxi\ lightcurve.
For the \Fexii\ lightcurve, the fit essentially returns the uniform prior on acceleration, so we do not show the corresponding posterior for $v_{20}$.
The same is true for all fits to the 2017 epochs.

For the March 2017 epoch, lightcurves with fits are shown in Figure~\ref{fig:blc} and the posteriors on EE mass are shown in Figure~\ref{fig:bmasses}.
For the August 2017 epoch, lightcurves with fits are shown in Figure~\ref{fig:clc} and the posteriors on EE mass are shown in Figure~\ref{fig:cmasses}.
Corner plots of fit parameters for these epochs are available in Appendix~\ref{app:morefits}.
Nonzero dimming depths produce better fits in 3 of 4 cases, but the Bayes Factors for nonzero versus zero depth range from 0.2-0.8, favoring the null hypothesis of no dimming.

\begin{figure*}
\includegraphics{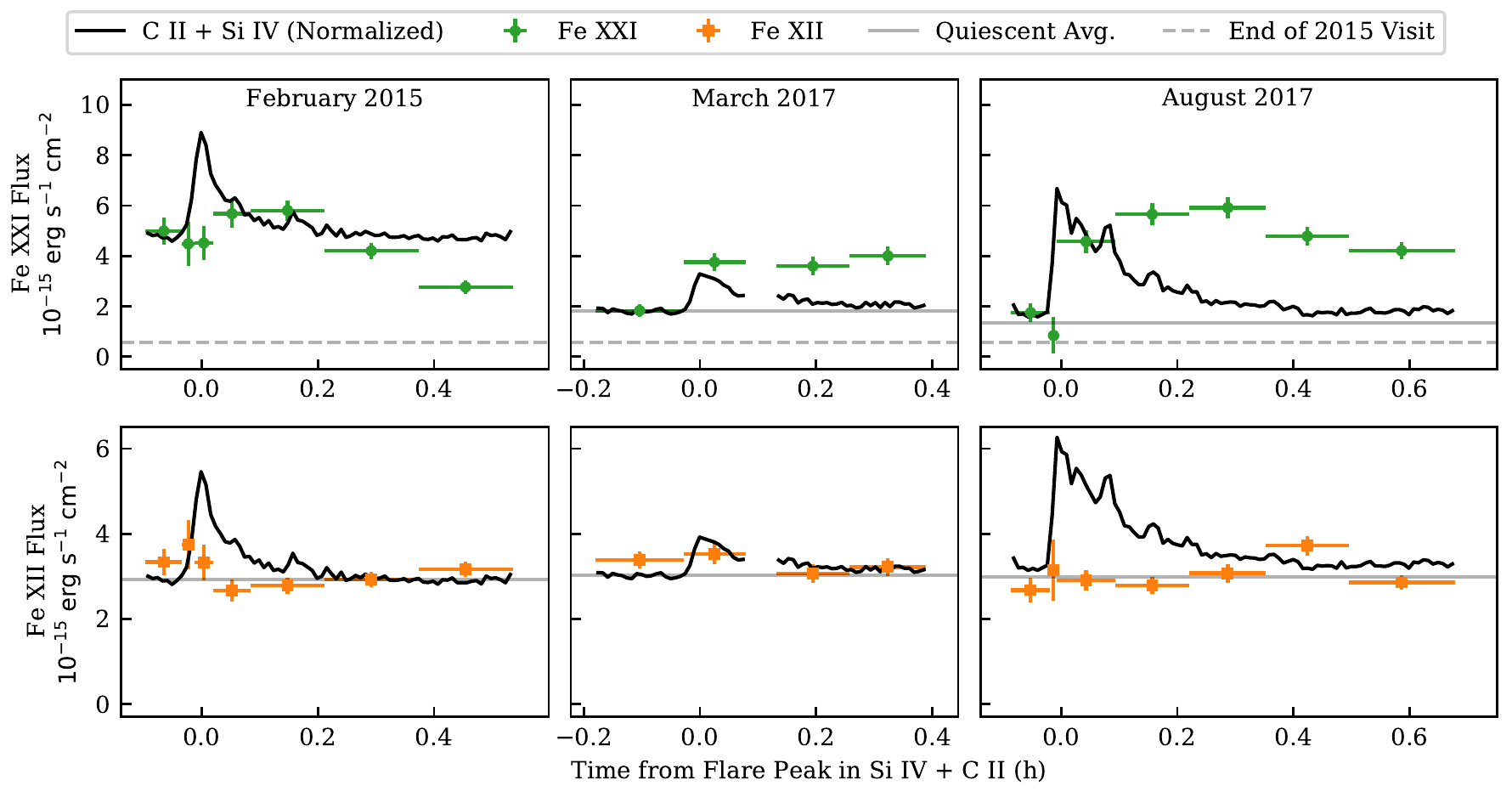}
\caption{
Zoom in on the flares from each observing epoch plotted in absolute flux on the same scale to enable a comparison between epochs.
The solid gray lines mark the average flux outside of flare.
The dashed gray lines in the \Fexxi\ plots mark the average flux from the last exposure of the February 2015 visit (i.e., following the candidate dimming).
\Fexxi\ fluxes were computed using the ``line ratio'' method (green points in Figure \ref{fig:lc}) and \Fexii\ fluxes are from direct integration (orange points in Figure \ref{fig:lc}).
Note in the top row that the \Fexxi\ flux at the start of the February 2015 visit is elevated well above other visits (gray lines), yet declines to a value (dashed line) well below the out-of-flare average for the other visits. }
\label{fig:zoom}
\end{figure*}

\section{Discussion}
\label{sec:discussion}

\subsection{The February 2015 Event}
\label{sec:event}
Following the February 2015 flare, flux from the \Fexxi\ line nearly disappears, dropping to roughly 20\% of its pre-flare level (Figure \ref{fig:lc}).
Could a CME be responsible?
An 80\% dimming is unheard of on the Sun, where the largest dimmings rarely exceed 10\% in magnitude \citep{mason19}.
Solar coronal dimmings also typically occur in cooler lines, near the $\sim$1~MK peak of the solar EMD, such as \ion{Fe}{12} \citep{mason14}.
However, \ee's EMD peaks at hotter temperatures, and in 2015 exhibited a plateau out to 10~MK (Figure~\ref{eq:em}), suggesting that coronal dimming could be expected in hotter, higher ionization iron lines in comparison to the Sun, perhaps even \Fexxi.
\edit1{Simulations by \cite{jin20} indicate that dimming on more active stars could be shifted to higher-temperature lines relative to the Sun.}

The 2015 flare on \ee\ began only 3.8~min after the start of the observations.
This proximity to the start of the exposure suggests an alternate explanation for the \Fexxi\ dimming: that the initial ``pre-flare'' \Fexxi\ flux is elevated and its subsequent decay is simply a return to its original quiescent value rather than a dimming.
An elevated flux could be due to an early response to the observed flare, preceding that of \Siiv\ and \Cii, or due to a larger, unseen flare that occurred prior to the start of the observing visit.
To help evaluate the 2015 event, Figure \ref{fig:zoom} provides a closer look at each of the observed flares.

\edit1{Points c}hallenging the CME explanation are that the peak of the flare in \Fexxi\ is subdued and its pre-flare value is significantly higher than in the 2017 epochs.
\Fexxi\ flux increases 1.2-1.5$\times$ during the 2015 flare versus 2-4$\times$ in the 2017 flares.
The onset of dimming offsets this peak, but only mildly.
Accounting for it only boosts the flare increase to $1.7\pm0.4$.

Before the flare, the \Fexxi\ flux of the 2015 observations is elevated by about 3$\times$ relative to 2017 (Figure \ref{fig:zoom}).
This elevated flux is probably not an early increase due to the observed flare.
The rise in \Cii\ + \Siiv\ flux clearly begins later in the exposure, and \Fexxi\ flux does not lead increases in this emission in later epochs.
Further, on other stars transition region lines appear to brighten nearly simultaneous with the particle acceleration and heating from a flare \citep[e.g., ][]{macgregor21}.
What is instead possible, and cannot be ruled out by the brief 3.8~min baseline, is that a larger flare occurred prior to the start of the 2015 observations.
In this scenario, the elevated \Fexxi\ emission and subsequent dimming are just the ongoing decay of the unseen flare.
Although the \Cii\ + \Siiv\ emission does not appear elevated at the start of the observations, this is consistent with their more rapid decay following flares, clearly visible in the 2017 observations.

A hidden flare is statistically plausible.
The probability of a hidden flare is roughly $\mu \tau / \alpha$, where $\mu$ is the rate of flares producing peak fluxes at or above the observed level, $\tau$ is the flare decay timescale, and $\alpha$ is the power-law index of the (cumulative) flare frequency distribution (see Appendix~\ref{app:preflare} for a derivation).
We estimated $\mu$ using the two largest flares in the other epochs of observation.
Their \Fexxi\ flux peaks near the starting level of the 2015 epoch, meaning that flares exceeding this threshold occur at an approximate rate of 0.1~ks$^{-1}$.
We estimated $\tau$ by fitting an exponential decay to the 2015 event (i.e., assuming the dimming is a flare decay), obtaining a value near 7~ks.
We assumed $\alpha$ is similar to estimates made using solar and stellar events, which generally yield values between 0.5 and 1.5 \citep{aschwanden21}.
The resulting probability that the observed decay is actually due to an unseen flare is greater than 10\%.

What is odd is that, by the end of the 2015 observations, the \Fexxi\ flux has settled to a value well below the pre- or post-flare flux of the 2017 observing epochs, by about a factor of 3.
This means that neither the pre- nor post-flare \Fexxi\ flux in 2015 is consistent with that of the later epochs.
The dimming hypothesis requires that the baseline flux was 3$\times$ higher than in later epochs, whereas the hidden flare hypothesis requires that the baseline flux was 3$\times$ lower.

All observing epochs occur near minima in \ee's $2.92$~yr activity cycle \citep{coffaro20}.
X-ray observations by XMM-Newton simultaneous with the 2015 visit measured a broadband flux of $1.1\sn{-11}$~\fluxcgs, lower than measurements of $1.5\sn{-11}$ and $1.4\sn{-11}$~\fluxcgs\ that occurred -50 and +8 days from the March 2017 and August 2017 HST visits.
The average \Fexii\ emission varied little across all three visits (Figure \ref{fig:zoom}).
On the Sun, activity cycles affect emission from hotter lines \edit1{more}, as indicated by an increasing variability in the solar EMD with increasing temperature.
However, this does not seem to be the case on \ee\ \citep[e.g.,][]{coffaro20}.

The 2015 X-ray data begin just after the peak of the observed flare and mimic the decline of \Fexxi.
The ratio between the peak and final values is $\approx$1.5, much less than the drop in \Fexxi\ emission \citep[Figure 9]{loyd18}.
Like \Fexxi, this drop in X-ray flux could also result from either a flare decay or coronal dimming.

% # Time between start of obs and start of flare
% t0 = np.min(p.obs_times)
% t0 = t0*u.s + p.time_datum
% (flares.a.start - t0.jd)*60*24
%
% peak flux came from mousing over the python plot (plots.lightcurves())
%
% # pre flare flux and flux ratios
% for lbl in 'abc':
%     fe21 = pubdata.lightcurves[lbl]['Fe21'].meta['r_normflux']
%     fe12 = pubdata.lightcurves[lbl]['Fe12-1'].meta['d_normflux']
%     ratio = fe21/fe12
%     print('Fe XXI: {:.1e}'.format(fe21))
%     print('Fe XII: {:.1e}'.format(fe12))
%     print('Ratio: {:.1f}'.format(ratio))
% 5/1.74
% 5/1.48
%
% post flare flux just comes from knowing that the max likelihood dimming is % ~90%

If the CME explanation is correct, it could indicate that the 2015 event resulted from the ejection of plasma from a single dominant active region, emitting, outside of flares, 3$\times$ the flux of active regions at other epochs.
In this picture, the magnetic reconnection that produced the flare in \Siiv\ and \Cii\ ejected the bulk of the \Fexxi-emitting plasma in that region, producing the deep \Fexxi\ dimming.
Meanwhile, either the \Fexii-emitting plasma in that region was preserved, or what plasma was ejected made up a much smaller fraction of the star's disk-integrated \Fexii\ emission.

A third possibility for the dramatic drop in \Fexxi\ emission is that a dominant active region rotated out of sight over the stellar limb during the observation.
However, the 11~d rotation period of \ee\ makes this unlikely, as the duration of the visit occupies only about 2.5\% of this period.
A bright region receding across the limb would have a redshift below 2.5~\kms \citep{valenti05}, too low to test this theory given the instrument accuracy of 7.5~\kms \citep{plesha19}.

That a single emission line could dim by as much as 80\% following a CME is made more plausible by the recent work of \cite{veronig21}.
That work discovered dimmings as high as 56\% in broadband X-ray emission from several stars and attributed those dimmings to CMEs.
A 56\% dimming in the bands they analyzed, which blend lines and continuum, could be deeper in an isolated line.

Like so many past CME candidates, the real cause of the 2015 dimming is simply inconclusive.
However, this event hints at the potential power of searching for coronal dimming with FUV observations.
Dedicated observations of this or other bright, active stars could capture many more flares with better pre-flare baselines and firmly establish whether near total dimmings of the \Fexxi\ line do indeed occur, or if the 2015 event was more likely to be the result of an unseen flare.

The enticing \Fexxi\ signal should not overshadow what we consider the most valuable result of this analysis: actual constraints (in this case, upper limits) on the masses of emitting material that could have been ejected in association with flares on \ee.
This is the subject of the next section.

\subsection{Mass Estimates: Context and Limitations}
\label{sec:caveats}
On the Sun, the strongest flares in the 3 epochs of observation would likely each have been accompanied by a CME.
The threshold above which flares on the Sun are always accompanied by CMEs is a peak soft X-ray flux exceeding roughly $3\sn{-4}$~W~m$^{-2}$ \citep{harra16}.
The peak soft X-ray fluxes of the 3 \ee\ flares were in the range of $10^{-4}$ - $10^{-2}$~W~m$^{-2}$ (GOES class X to X100) according to the empirical relationship between peak \Siiv~1393,1402~\AA\ and soft X-ray flux established by \cite{youngblood17}.
\edit1{This range is driven by uncertainty in the empirical relationships for predicting the soft X-ray flux from peak \Siiv\ flux at 1~AU.
The peak \Siiv\ fluxes at 1~AU of the 3 observed flares were 0.39, 0.25, and 0.44~erg~\pers~\percmsq.
The solar flare-CME scaling of \cite{aarnio12} indicates that flares of this magnitude on the Sun could be associated with CMEs ranging from $10^{15} - 10^{18}$~g.
This mass range would correspond to dimming depths ranging from undetectable to physically impossible (i.e., $>100$\%, assuming a single emission line probed the full CME mass).}

\begin{figure}
\includegraphics{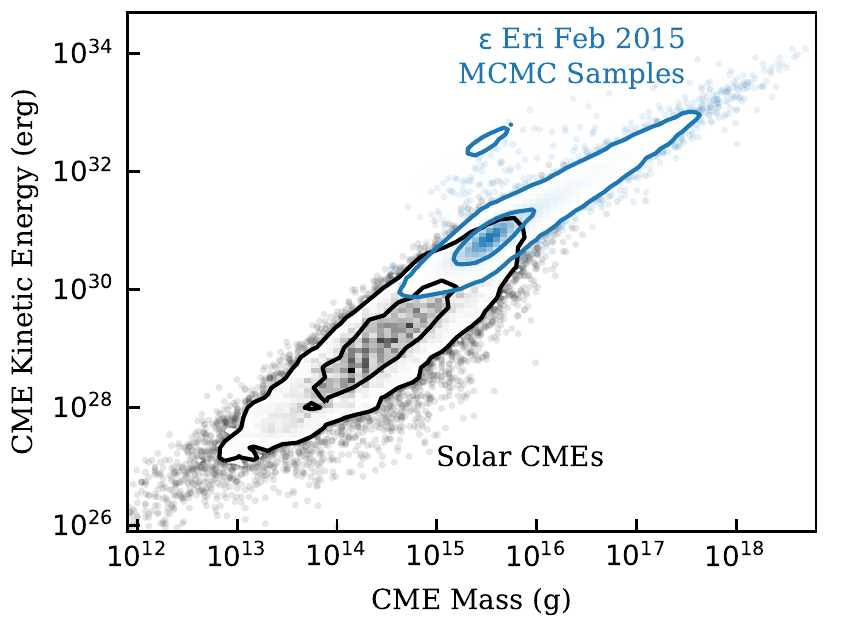}
\caption{
\edit1{Solar context for the February 2015 event if interpreted as a CME.
Black contours give the 1-$\sigma$ and 2-$\sigma$ bounds of solar CMEs.
Blue contours give the 1-$\sigma$ and 2-$\sigma$ bounds of the MCMC samples for the \Fexxi+\Fexii\ mass and kinetic energy of the February 2015 event on \ee.
The February 2015 event is consistent with solar CMEs, albeit on the high end of mass and energy.}
}
\label{fig:mvke}
\end{figure}

\edit1{If the February 2015 event was caused by a CME, we estimate an EE mass of $10^{15.2_{-1.0}^{+1.2}}$~g for the \Fexxi\ dimming and $10^{15.3\pm0.3}$~g for the \Fexii\ dimming (though the \Fexii\ dimming is statistically insignificant; see Section \ref{sec:analysis}).
% path = rc.path_mass_chain('a', 'fe12-1')
% chain = db.load_pickle(path)
% errs = utils.mcmc_errors(chain)
% np.log10(errs[0])
% np.log10(errs[0]) - np.log10(errs[0] - errs[1])
% np.log10(errs[0]) - np.log10(errs[0] + errs[2])
% path = rc.path_mass_chain('a', 'fe21')
% chain = db.load_pickle(path)
% utils.mcmc_errors(np.log10(chain))
The kinetic energy of this event, estimated by summing the \Fexxi\ and \Fexii\ masses and taking the velocity to be the $v_{20}$ values from the \Fexxi\ dimming, is $10^{30.8_{-0.3}^{+1.1}}$~erg.
% KE is output as part of the m_v_ke plot
Figure \ref{fig:mvke} plots the confidence interval of these estimates from the MCMC samples in the context of the population of solar CMEs from the SOHO/LASCO catalog \citep{gopalswamy09} through January 2021 (excluding those with data quality flags).
In that plot, the second mode of MCMC samples above the main mode correspond to the sampler exploring fits near the upper limit of the prior on CME acceleration that effectively amount to a step function in dimming.}

The dimming nondetections related to the 2017 flares yield \Fexii\ EE masses of $< 10^{15.5}$~g.
The \Fexxi\ EE masses are less constraining, at $< 10^{16.3}$~g.
% for lbl in 'bc':
%     for line in ['fe12-1', 'fe21']:
%         path = rc.path_mass_chain(lbl, line)
%         chain = db.load_pickle(path)
%         m = np.log10(np.percentile(chain, 95))
%         print('{} {} {:.1f}'.format(lbl, line, m))

In the context of solar CMEs, $10^{15}$-$10^{16}$~g CMEs are unremarkable events (1-2 orders of magnitude below the largest observed) and occur once every few days to every few months (\edit1{Figure~\ref{fig:mvke}} and Section \ref{sec:massloss}).
One might expect the EE masses to be \edit1{well} above the upper limit of solar CME masses given that \edit1{\ee\ has a larger peak emission measure (Figure~\ref{fig:emd}),} the \Fexxi\ dimming is extreme, and the \Fexii\ dimming large in comparison to solar events.
The relatively low EE masses in this context likely result from a combination of factors: plasma densities that are 2-3 orders of magnitude above the solar corona \citep{fludra99}, the exclusion of plasma beyond the range of plasma temperatures \edit1{and densities} probed by the lines, \edit1{and the fact that solar CME masses grow as they leave the Sun prior to reaching the mass recorded in the SOHO/LASCO catalog \citep{gopalswamy09}.}

The mass posterior in Figure \ref{fig:masses} exhibits a cliff at its low-mass end.
This is due to a near-linear decline in the emissivity, $G(T,n)$, above $n=10^{13}$~\percmcb\ as collisional quenching of the (forbidden) \Fexxi~1354~\AA\ radiative transition sets in.
Because $m \propto (nG(T,n))^{-1}$ (Eqn. \ref{eq:m}), an increase in $n$ in this regime is effectively cancelled by a drop in $G(T,n)$, yielding a constant mass and producing the apparent lower limit in the mass posterior of Figure \ref{fig:masses}.

Dimming analyses of the sort we present might be the only means to probe the allowable masses of events associated with any prominent stellar flare.
However, it is important that we, as a community, recognize the assumptions and limitations of the method as we have presented it.
We list those of which we are aware below.
Many of these can likely be improved upon as the analysis techniques and observing strategies mature.

\begin{itemize}
    \item The method is blind to plasma beyond the narrow range of formation temperatures of the analyzed line(s).
    The mass of plasma with temperatures in the gaps will be missing from the estimated total.
    \item \edit1{The method is also blind to low-density, poorly-emitting plasma (Appendix \ref{app:density}).
    As a result, it likely probes mass in and around the erupting filament, but not mass above that is swept into the CME.}
    \item The method assumes that emission from the non-ejected plasma remains at the pre-ejection level.
    Changes in this baseline will affect the inferred dimming depth.
    \item The method equates the density of the ejected plasma to a ``disk-averaged'' density inferred from disk-integrated observations.
    If the ejected plasma is overdense relative to the disk-average, the retrieved EE mass will overestimate the true EE mass.
    \item Density estimates from observations simultaneous with the dimming observations are not always possible.
    Changes in density between these epochs will affect the EE mass.
    \item The method assumes the emission in an analyzed line is dominated by plasma at at a temperature that maximizes that line's emissivity.
    If the plasma contributing most of this emission has a non-peak temperature, the retrieved EE mass will be an underestimate.
    \item The method assumes no plasma remains within the dimming region.
    \edit1{If only a portion of the plasma is ejected, then the dimming region must be fractionally larger to account for the observed drop in emission. The net effect is that the ejected mass could drop to no less than half the original estimate (Section \ref{sec:method:mass}).}
    \item \edit1{The method assumes the CME has a constant mass.
    The mass of several solar CMEs has been observed to grow with time due both to mass added from behind and accumulated from ahead of the CME \citep[e.g.,][]{deforest13,veronig19}.}
    For reference, \cite{veronig19} reported a 5$\times$ increase in the mass of a solar CME observed on 2012 September 27.
\end{itemize}

EE mass estimates are subject to statistical uncertainties as well.
In the analysis of the \ee\ observations, the \Fexxi\ EE mass estimate has a range of 2 orders of magnitude because the density measurement we adopted from the literature has an equivalent range.
In contrast, uncertainty in both density and dimming depth contribute to the uncertainty on the \Fexii\ EE mass for that event.

The assumed emissivity also introduces an error, though it is negligible in the \ee\ analysis.
Figure~\ref{fig:emd} shows that the peak of the contribution functions is about 0.1~dex in temperature off from the peak of the line emissivities.
Adopting the peak of the contribution function to estimate $G(T,n)$ would increase the EE mass by about 30\%.

The most important point when it comes to temperature is the hidden mass.
For \ee, plasma from $\log(T)$ of 6.4 to 6.8 is effectively hidden (Figure~\ref{fig:emd}): plasma within this range will emit less than 10\% of the observed \Fexxi\ and \Fexii\ flux.
An ejection of plasma at these temperatures would have little affect on the observed \Fexii\ and \Fexxi\ emission.

MHD simulations of CMEs from active stars will help to determine how well observables like coronal dimming and Type-II radio bursts on \ee\ and other active stars can be expected to constrain CME properties.
\cite{o22} simulated Type-II radio burst signals from simulated CMEs on \ee\ and found that signals would remain in the LOFAR band for 20-30~min following an event.
This could make \ee\ a great candidate for joint radio and FUV observations to cross validate CME signals.
We are currently adapting the Alfven Wave Solar Model (AWSoM) for simulations of CMEs and dimming signals on \ee\ and will describe the results in a follow-on publication.
This will add additional context to the 2015 and 2017 HST observations.

% from Kevin: It would be cool to see if Meng’s models can predict the integrated X-ray dimming based on the observed Fe XXI depths (as well as the “more traditional” solar-dimming lines Fe IX, X, XI, XII, etc…like how big would this thing be if we observed it on the Sun?).

\subsection{CME Mass Loss Rate}
\label{sec:massloss}
The three observed flares and upper limits on the ejected emitting mass provide a basis to explore mass and angular momentum loss from \ee\ due to CMEs.

\begin{figure}
\includegraphics{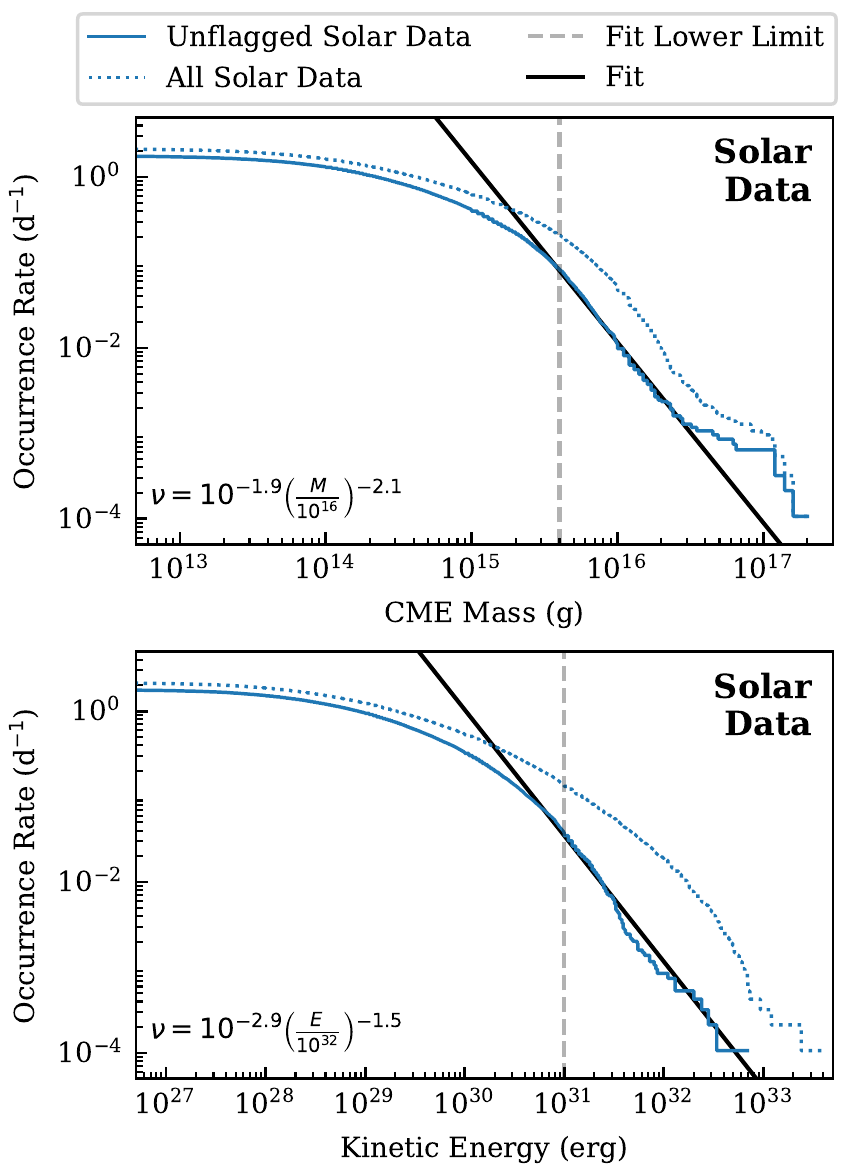}
\caption{
Solar CME masses and kinetic energies from the SOHO LASCO catalog \citep{gopalswamy09} through January 2021.
Power-law fits to data above a threshold (selected by eye) are also shown and the equation of the fit provided in the lower left of each panel.
The mass and energy estimates of many events are flagged as unreliable.
We fit only the unflagged events, but plot curves both for all events and unflagged events.
There was no mass or energy estimate for 41\% of events in the LASCO catalog.
% number printed when you run plots.solar_cme_ffds()
}
\label{fig:cme_fd}
\end{figure}

A challenge to estimating CME-driven mass loss is that events spanning orders of magnitude in mass and occurrence rate could contribute.
This is certainly true on the Sun.
Figure \ref{fig:cme_fd} plots the distribution of observed CME masses from the SOHO LASCO catalog.
Summing the masses of all events with mass estimates (including those with quality flags) yields a CME mass loss rate of $6\sn{-16}$ \Msun\ \peryr.
Small events contribute the most to CME mass loss on the Sun.
Events with masses less than 3\% of the largest observed (i.e. $<5\sn{15}$~g) contribute more than 50\% of the summed mass of all events in the LASCO catalog.

Assuming a power-law distribution of CME masses can enable an estimate of cumulative CME mass loss incorporating events over a range of masses.
More sophisticated treatments are also possible \citep{odert17}.
Portions of the observed solar CME mass distribution follow a power law reasonably well (Figure~\ref{fig:cme_fd}).
For a power law of the form
\begin{equation}
    \nu(>m) = \mu (m/\mMref)^{\alpha},
    \label{eq:ffd}
\end{equation}
where $\nu$ gives the rate of events with mass $>m$ and $\mu$ is the rate of events with mass $>\mMref$.
Events within the range $m_\mathrm{min}$ and $m_\mathrm{max}$ account for a cumulative mass loss of
\begin{equation}
    \mMdot = \frac{\alpha \mu \mMref}{\alpha + 1} \left[ \left( \frac{m_\mathrm{max}}{\mMref} \right)^{\alpha + 1} - \left( \frac{m_\mathrm{min}}{\mMref} \right)^{\alpha+1} \right].
    \label{eq:mdot}
\end{equation}
For $\alpha \approx -2$, as on the Sun, small events dominate and \Mdot\ is not sensitive to $m_\mathrm{max}$.
Integrating the power law from where it begins to deviate significantly at $m_\mathrm{min} = 2\sn{15}$~g to $m_\mathrm{max} \rightarrow \infty$ yields a mass loss rate within a factor of a few of the sum of the cataloged solar CMEs.

For \ee, taking $m_\mathrm{ref} < 3\sn{15}$~g based on dimming fits to the 3 prominent flares within the 10.6~h of observations ($\mu \approx 7$~d$^{-1}$), and adopting solar values for the remaining parameters, implies $\mMdot < 10^{-14}$~\Msun~\peryr, or $\mMdot < 0.6$~\Mdotsun\ (where we take \Mdotsun\ to be $2\sn{-14}$~\Msun~\peryr\ for consistency with \citealt{wood05}).
We emphasize that this estimate is confined only to \Fexii-emitting plasma and only to flare-associated CMEs.
The total mass loss rate of \ee, incorporating CMEs and the stellar wind, has been estimated from the astrospheric absorption method to be $\mMdot = 30$~\Mdotsun\ \citep{wood05}.
Our results suggest that CMEs contribute only negligibly to \ee's total mass loss, similar to the present-day Sun.
This conflicts with the MHD simulations of \cite{o22} that, coupled with flare statistics, indicate CMEs could shed mass from \ee\ at up to half the rate of the stellar wind.
For any statement about \ee's mass loss rate to progress beyond conjecture, constraints across a wide range of CME rates and masses are needed.
% a = 2.1
% mmin = 2e15*u.g
% mref = 1e16*u.g
% mu = 10**-1.9/u.d
% mdot = mu*a/(a+1)*mref**-a*mmin**(a+1)
% mdot.to('Msun yr-1')
% mu = 3/10.6/u.h
% mref = 3e15*u.g
% mdot = mu*a/(a+1)*mref**-a*mmin**(a+1)
% mdot.to('Msun yr-1')
% ratio = mdot/Mdotsun
% ratio.to('')

Observationally calibrated scalings between CME mass and flare energy provide another path to constraining the CME mass loss rate \citep{odert17}.
For the Sun, \cite{aarnio13} measured $\mMcme \propto E_\mathrm{flare}^{0.63}$.
The sub-linear relationship is consistent with solar CME masses following a steeper distribution ($\alpha \approx -2$) than solar flare energies ($\alpha \approx -1$; \citealt{aschwanden16}).\footnote{Note that in \cite{aschwanden16} $\alpha$ is defined for differential distributions, whereas we define it for a cumulative distribution. Values differ by 1 as a result.}
Empirical flare-CME scalings for stars could allow future work to fold long-baseline observations of a star's flares from missions like \textit{TESS} into determining the cumulative effects of stellar CMEs.
Calibrating such a scaling for \ee\ or other stars will require measured CME masses across a range of flare energies, so is not possible with the present data.

\subsection{Confined CMEs}
\label{sec:confinement}

\edit1{\cite{alvarado18,alvarado19} modeled CMEs triggered beneath global dipolar magnetic fields ranging from 75-1400~G representative of active stars \citep{donati09,reiners22} and found such global fields could prevent incipient CMEs from escaping.
Strong magnetic fields are typical of low mass stars in particular.
The relationship between average field strength and Rossby numbers implies that, for M stars, even those that have spun down with age to rotation periods of 100~d ($Ro \approx 1$; \citealt{wright18}) will exhibit average field strengths of 100~G and above \citep{reiners22}.}

\edit1{Magnetic confinement} could explain why it has proven difficult to detect the exceptional CMEs one might expect around stars with exceptionally energetic flares \citep{drake13}.
Evidence of strong fields confining incipient CMEs has been observed on the Sun \citep{thalmann15}, and a reanalysis of purported stellar CME detections in the literature suggests these events, though they ultimately escaped, spent much of their kinetic energy breaking free of overlying fields \citep{moschou19}.

The simulations of \cite{alvarado19} produced a sustained dimming in a band at 171~\AA\ (dominated by \ion{Fe}{9} emission) regardless of whether the simulated event escaped or not.
This implies that coronal dimming on active stars could be a result of ``failed'' CMEs that do not actually escape the star.
\edit1{On the Sun,} flares lacking CMEs produced no dimming signature in the analysis of 42 flares, 9 without CMEs, by \cite{harra16}.
Six of these were from active region NOAA12912, where strong overlying fields likely stymied incipient CMEs \citep{thalmann15}.
\edit1{\cite{veronig21} also analyzed a sample of solar flares as a precursor to their search for coronal dimmings in stellar data.
Within the 64 solar flares they analyzed, they found that of the 16 events without an accompanying CME, 2 produced spurious dimming signatures.}
More work is needed to understand what factors result in a CME false-positive through dimming.
However, even if coronal dimming can result from eruptions that do not ultimately escape, it remains a valuable observable.
In this case, the absence of dimming would indicate no eruption, and certainly no CME, occurred.
Further, rates of confined eruptions, potentially probed by dimming, could provide some sense of the rate of more powerful eruptions that do escape, even if they are too infrequent to easily observe.

\edit1{For \ee, CME confinement is possible, but uncertain.
The maximum strength of \ee's (dominantly poloidal) surface magnetic field was 20-33~G near the 2015 observations \citep{jeffers17}, versus 8~G for analogous observations of the Sun in 2013 \citep{vidotto16}.
At the 75~G field strengths modeled by \cite{alvarado18}, only eruptions associated with flares of GOES class X30 or greater could escape confinement.
This falls within the range of uncertainty on the GOES class of the \ee\ flares we analyzed, making it unclear whether associated eruptions should have been capable of breaking free of an overlying field.
The February 2015 event, if a CME, had a kinetic energy consistent with the distribution of kinetic energies of solar CMEs for events of similar mass (Figure~\ref{fig:mvke}), in contrast with the more extreme events analyzed by \citep{moschou19}.
Note, however, that the $v_{20}$ value we used to estimate the 2015 CME candidate's kinetic energy might not be sensitive to confinement, since this could slow the CME after the initial acceleration probed by dimming and used to estimate $v_{20}$.
}

\subsection{Considerations for Future Coronal Dimming Observations}
\label{sec:hst}

The ideal coronal dimming observations would incorporate emission lines tracing a broad range of formation temperatures, with emphasis on those near the peak of the stellar EMD.
The benefit of this coverage would be minimizing the potential for ejected mass to be hidden from a dimming analysis.
CME mass constraints for \ee\ in particular would benefit from observations of lines formed between $\log{T}$ of 6.4 to 6.8.

With sufficient coverage of coronal emission lines and sufficient SNR, it would be possible to construct EMDs for dimming events themselves.
Coupled with measurements of plasma density from lines tracing a similar range of temperatures, one could construct a distribution of mass versus temperature, $dm/dT$, and integrate this distribution to obtain a total mass.
Future work could test such an approach on observations of solar CMEs.

Observations of coronal (irradiance) dimming on the Sun rely on EUV emission lines, particularly those near 200~\AA, such as observed by SDO/EVE \citep{mason16}.
From 100-350~\AA, the EUV band contains many bright Fe lines spanning a range of ionization states and formation temperatures.
No astrophysical observatories currently operate that can access 100-350~\AA\ emission.
The former Extreme Ultraviolet Explorer (EUVE), which operated from 1992 to 2001, could observe this range.
\cite{veronig21} included these archival data in their comprehensive search for coronal dimmings, identifying a single candidate dimming from the EUVE data on AB~Dor.

X-ray observations are dominated by coronal emission.
Effective areas are 1-3 orders of magnitude below those of HST/COS, but because the emission is dominantly coronal, wide bands can be integrated to achieve higher SNR.
For the star EK~Dra, which has a nearly identical flux in \Fexxi~1354~\AA\ emission as \ee\ \citep{ayres10}, the analysis by \cite{veronig21} yielded an SNR of $\approx130/\sqrt{\mathrm{hr}}$ in a 0.2-12~keV band and $\approx60/\sqrt{\mathrm{hr}}$ in a narrower 1-2~keV band.
In comparison, median SNRs were 17/$\sqrt{\mathrm{hr}}$ in \Fexxi\ and 40/$\sqrt{\mathrm{hr}}$ in \Fexii\ for the 2017 HST COS observations of \ee.
The higher SNR of broadband X-ray searches comes at the cost of being able to isolate emission lines in order to enable mass estimates through the process outlined in Section \ref{sec:method}.
FUV spectroscopy with HST also offers simultaneous measurements of 10-100$\times$ brighter chromospheric (e.g., \Ci) and transition region (e.g., \Siiv) lines that can be used to closely trace flares.

For dimming observations seeking mass constraints, an important consideration is how to estimate plasma densities.
In the FUV, observing both the 1242~\AA\ and 1349~\AA\ components of \Fexii\ enables a simultaneous density measurement of the \Fexii-emitting plasma.
Unfortunately, no additional bright components of \Fexxi\ are available in the FUV that could enable a direct measurement of its source plasma density.
Coupling HST observations with contemporaneous X-ray observations of a density-sensitive doublet like \ion{Ne}{9} \citep{ness02} will be important to obtain mass estimates of ejected \Fexxi-emitting plasma.

When using HST for FUV measurements of coronal emission, COS provides the best combination of sensitivity and wavelength coverage.
However, for many active stars, especially M dwarfs, the potential for flares to violate detector count rate limits is a concern.
Using the Space Telescope Imaging Spectrograph (STIS) enables the observation of brighter targets in exchange for lower sensitivity.
The STIS G140M mode provides a throughput of 1.7\% at 1350~\AA\ \citep{branton21} versus COS G130M's 4.0\% \citep{hirschauer21}.

\section{Summary}
\label{sec:summary}
We have adapted the technique of using coronal dimming to estimate the mass and velocity of solar CMEs to stellar observations, providing generalized formulae for order-of-magnitude estimates.
A strength of this method is that it can be applied to nondetections, providing limits on the maximum mass of emitting material that was ejected.

FUV spectroscopy enables dimming measurements using the \Fexii~1242 and 1349~\AA\ lines that trace $\sim$1~MK plasma as well as the \Fexxi~1354~\AA\ line that traces $\sim$10~MK plasma.
We conducted a dimming analysis of these lines in archival observations of the young (200-800~Myr) solar-type star \ee.
Each of three separate epochs of these observations contained a prominent flare with a peak \Siiv~1393,1402~\AA\ flux that, on the Sun, would correspond to strong flares always accompanied by CMEs.

Following a flare in February 2015, \Fexxi\ emission dropped 81$\pm$5\% below its pre-flare level and \Fexii\ emission dropped by 16$\pm$8\%.
% chain = db.load_pickle(rc.path_fit_chain('a', 'fe21'))
% chain = chain['chain']
% from mypy import pdfutils as pdf
% utils.mcmc_errors(chain[:,2])
% chain = db.load_pickle(rc.path_fit_chain('a', 'fe12-1'))
% chain = chain['chain']
% utils.mcmc_errors(chain[:,2])
If caused by a CME, the dimmings imply that $10^{15.2_{-1.0}^{+1.2}}$~g of 10~MK plasma was ejected, possibly accompanied by $10^{15.3\pm0.3}$~g of 1~MK plasma.
However, a viable alternate explanation for this event is that it is the continued decay of a flare that occurred prior to the start of the observations.
The initial baseline is too short to distinguish these scenarios, and \ee's high flare rate makes a hidden flare statistically plausible.

Two other flares at later observing epochs did not exhibit significant \Fexii\ or \Fexxi\ dimming.
For these events, we found upper limits of a few $\times10^{15}$~g in ejected 1~MK (\Fexii-emitting) plasma and a few $\times10^{16}$~g in ejected 10~MK (\Fexxi-emitting) plasma. Ejections of this or greater magnitude must occur less than a few times a day on \ee.
On the Sun, CMEs with total masses of $10^{15}$~g happen once every few days.
With some assumptions, we suggest the mass loss rate of 1~MK plasma due to flare-associated CMEs is $<0.6$~\Mdotsun, in comparison to the star's measured mass loss rate (CMEs $+$ wind) of 30~\Mdotsun\ \citep{wood05}.

This work establishes the viability of coronal dimming as a method to detect stellar CMEs and constrain their masses even if not detected.
Combining dimming with other methods, such as Type-II radio bursts or Doppler shifts in line emission, could yield particularly strong evidence of stellar CMEs.
Identifying and maturing reliable methods to constrain stellar CMEs is essential if we wish to accurately assess the role of CMEs in stellar and planetary evolution.
\\
\\
\\
The authors thank the anonymous reviewer for their constructive critique that substantially improved the quality of this work. 
ROPL conducted this research under program HST-AR-15803.
Support for program HST-AR-15803 was provided by NASA through a grant from the Space Telescope Science Institute, which is operated by the Associations of Universities for Research in Astronomy, Incorporated, under NASA contract NAS5- 26555.
This research is based on observations made with the NASA/ESA Hubble Space Telescope obtained from the Space Telescope Science Institute, which is operated by the Association of Universities for Research in Astronomy, Inc., under NASA contract NAS 5–26555. These observations are associated with program(s) 13650, 14909, and 15365.
CHIANTI is a collaborative project involving George Mason University, the University of Michigan (USA), University of Cambridge (UK) and NASA Goddard Space Flight Center (USA).
The solar CME catalog used in this work is generated and maintained at the CDAW Data Center by NASA and The Catholic University of America in cooperation with the Naval Research Laboratory. SOHO is a project of international cooperation between ESA and NASA.

\software{emcee \citep{foreman13}, CHIANTIpy (https://chiantipy.readthedocs.io), calCOS (https://github.com/spacetelescope/calcos)}

\facility{HST (COS)}

\appendix

\section{Derivation of the Probability of an Unseen Flare}
\label{app:preflare}
We provide here a derivation for estimating the probability that an unseen flare is responsible for an elevated flux at the beginning of an observation.
We consider only single flares because the probability of two or more flares all contributing in roughly equal proportion to the flux is small \citep{loyd18}.
Even in cases where flares overlap, one flare usually dominates.

We take the number of flares, $d^2N$, occurring over time $dt$ and that have a peak flux within $dF_p$ of $F_p$ to follow a power law of the form
\newcommand{\Fref}{F_\mathrm{ref}}
\begin{equation}
    d^2N = \frac{\alpha \mu}{\Fref}  (F_p/\Fref)^{-(\alpha+1)} dF_p dt
    \label{eq:dffd}
\end{equation}
The grouping of constants will yield a cleaner final product from this derivation.
It also yields a simple form for the cumulative distribution, giving the number of flares with peak flux $> F_p$ as
\begin{equation}
    dN(>F) = \mu (F_p/\Fref)^{-\alpha} dt.
    \label{eq:cffd}
\end{equation}
The normalization constant $\mu$ indicates the rate of flares with $F_p > \Fref$.
The use of $\Fref$ and $dt$ is uncommon when working with flare frequency distributions, but we include them for clarity.
The $dt$ indicates that the number of flares expected depends on the amount of time considered and the $\Fref$ mitigates confusion regarding units raised to arbitrary powers of $\alpha$.
Note that peak flux could be replaced with any value that quantifies the strength of a flare, such as energy.

As a second key assumption, we take flares to have a consistent time profile defined for any time following the peak of the flare as
\begin{equation}
F(t) = F_p e^{-t'/\tau}.
    \label{eq:decay}
\end{equation}
For convenience, we will set $t'$ such that the peak of the flare occurs at $t'=0$ and the observation at $t'=t$.
This implies that to produce the flux observed, $F_0$, a flare that peaked at an (unobserved) time $t$ before the start of the observations must have had a peak flux of
\begin{equation}
F_p = F_0 e^{t/\tau}.
    \label{eq:peak}
\end{equation}
In words, the farther back the flare occurred, the larger it must have been in order to explain the observed flux.

Combining these relationships, the expected number of a flares peaking within $dt$ of time $t$ prior to the observation is
\begin{equation}
    d^2N = \frac{\alpha \mu}{\Fref} e^{\alpha t/\tau} (F_0/\Fref)^{-(\alpha+1)} dt dF_0,
\end{equation}
where we have made the change of variables from $dF_p$ to $dF_0$ noting that $dF_p/dF_0 = e^{t/\tau}$.
Integrating over all possible times an unseen flare could have occurred ($-\infty < t \leq 0$), gives
\begin{equation}
    dN = \frac{\tau \mu}{\Fref} (F_0/\Fref)^{-(\alpha+1)} dF_0,
\end{equation}

The expected number of flares that could have produced a flux $> F_0$ at the time of observation is then
\begin{equation}
    N = \frac{\tau \mu}{\alpha} (F_0/\Fref)^{-\alpha}.
    \label{eq:unseen}
\end{equation}
Taking flares to follow Poisson statistics with respect to their occurrence over time, the likelihood that one or more events occurred is $p = 1 - e^{-N}$.
For small $N$, $p \approx N$.
Monte-Carlo experiments confirm the validity of Eqn.~\ref{eq:unseen}.

\section{Effects of Density Variations}
\label{app:density}

One of the key assumptions of this work is that the plasma is homogeneous in density.
A true plasma will vary in density due to gravity, magnetic topology, and other factors.
Here we explore the effect of this when estimating masses using the framework of Section~\ref{sec:method:mass}.
Recall the mass estimate is based on an observed flux decrement and density estimated from a line ratio.

The integral given in Eqn.~\ref{eq:em} can be rewritten
\begin{equation}
    F = \int_{n_a}^{n_b} n^2 G \frac{dV}{dn} dn,
    \label{eq:hF}
\end{equation}
where $n_a$ and $n_b$ give the lower and upper limits of the densities spanned by the overall volume of emitting mass.
We have set $n\equiv n_e = n_H$ and abbreviate $G(T,n_e)$ as $G$.
The mass of the plasma is
\begin{equation}
    m = \mu \int_{n_a}^{n_b} n \frac{dV}{dn} dn,
    \label{eq:hmass}
\end{equation}
where $\mu$ is the mean particle mass.

The quantity $dV/dn$ represents the distribution of mass across different densities, i.e., the volume of mass associated with a infinitesimal interval of density.
We will refer to $dV/dn$ as ``clumpiness.''

We explored power law clumpiness distributions to determine the effect on mass, i.e.,
\begin{equation}
    \frac{dV}{dn} = C n^\alpha,
\end{equation}
where $C$ is a scaling factor that sets the total volume (hence mass and emitted flux) of the plasma.
Both a hydrostatic volume (with constant scale height) and a Parker Wind (with constant outflow velocity) fall in the category of power-law clumpiness, with $\alpha = -1$ and -2.5 respectively.
An ideal hydrostatic atmosphere or Parker Wind would be better described as ``stratified'' than clumpy, but the $dV/dn$ distribution makes no assumptions about how the mass is organized, i.e., whether its density has smooth, monotonic spatial variations or it is truly clumpy.
Any $\alpha \leq -2$, including a Parker Wind, will yield a diverging (infinite) mass as $n_a \rightarrow 0$, whereas for $\alpha > -2$ the total mass is bounded.
Note the clumpiness could be any arbitrary function or measured distribution; we chose to investigate power laws simply because hydrostatic and Parker wind solutions are within the power law family.

Using this framework, we estimated the mass of heterogeneous-density volumes yielding the same flux ratios as the density-sensitive lines used in this work (\ion{Ne}{9} and \ion{Fe}{12}).
To do this, we chose several values for $\alpha$ and, for each, explored a grid of lower limits on density, $n_a$, down to several orders of magnitude below $n_\mathrm{ref}$, the constant-density value.
For each $n_a$, we numerically solved for the density upper limit, $n_b$, that would yield the appropriate flux ratio when Eqn.~\ref{eq:hF} was integrated for the density-sensitive lines.
With $n_b$ determined, we compared the mass, $m$, estimated from Eqn.~\ref{eq:hmass} to the mass assuming a constant density,
\begin{equation}
    m_\mathrm{ref} = \frac{\mu F}{n_\mathrm{ref}G},
\end{equation}
setting $F$ to be equal between the two estimates.

\begin{figure}
\includegraphics{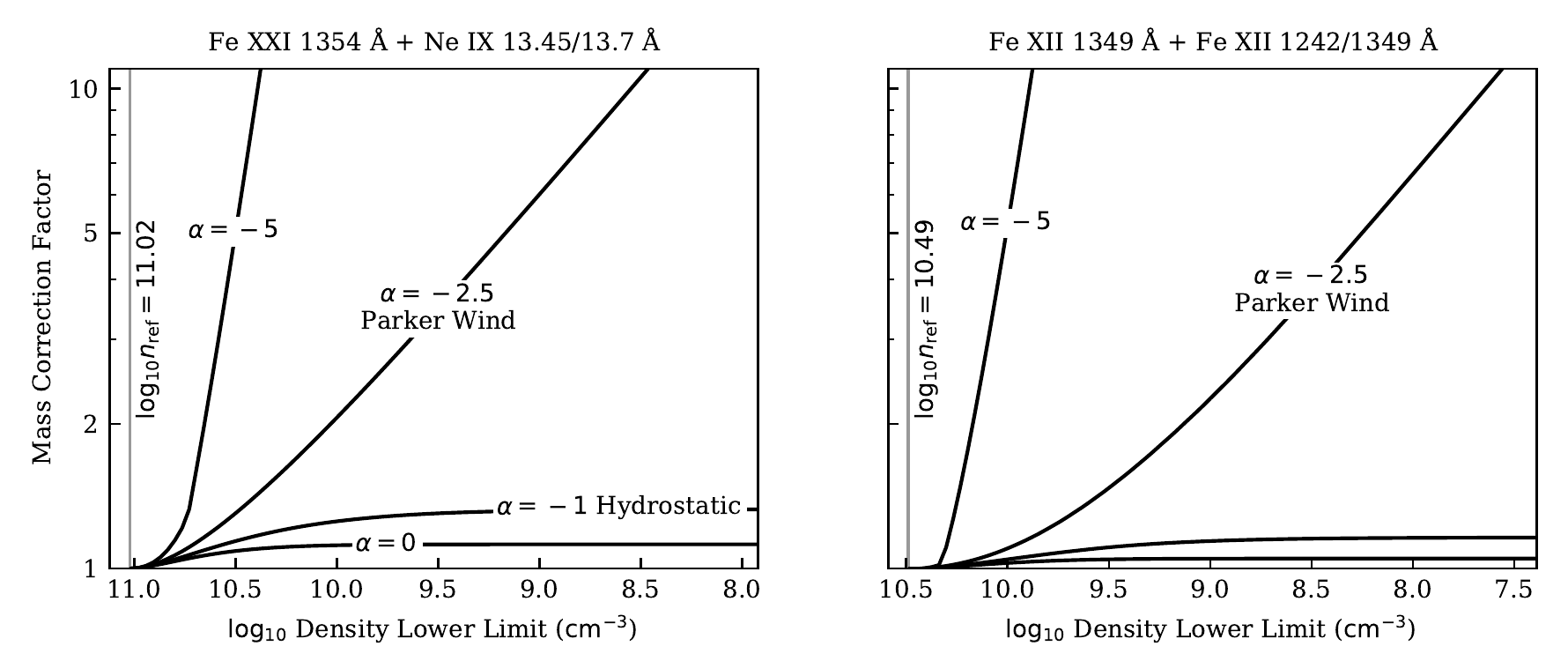}
\caption{
Mass increase when considering a heterogeneous-density distribution of the form $dV/dn \propto n^\alpha$ versus the homogeneous-density estimate adopted in the main body of this work.
Each curve adopts a different value for $\alpha$ and shows the additional mass as the volume is integrated to increasingly low densities while preserving the observed line ratio.
Large y-values mean significant mass could be present that is not accounted for if assuming a homogeneous density of $n_\mathrm{ref}$.
For $\alpha \leq -2$, significant low-density mass could be hidden from a dimming mass estimate.
}
\label{fig:masscorrection}
\end{figure}

Figure~\ref{fig:masscorrection} shows the resulting ``correction factors,'' $m/m_\mathrm{ref}$.
The hydrostatic case ($\alpha = -1$) yields small corrections of $<33$\% for our \Fexxi\ and \Fexii\ probes.
However, for $\alpha \leq -2$, including a Parker Wind ($\alpha = -2.5$), corrections can be well over an order of magnitude if the volume includes densities several orders of magnitude below the reference density.

This analysis demonstrates that mass well beyond the density suggested by the line flux ratio is essentially hidden.
In order to match the observed flux ratio, most of the flux must must come from mass near the density implied by that ratio.
For example, for the $\alpha=-2.5$ case, even if we integrated the density to a lower limit of 0 when computing fluxes, the upper limit needed to be extend only an extra 0.5~dex above $n_\mathrm{ref}$ to compensate for the emission of the lower-density plasma and preserve the observed line flux ratio.
This is in spite of the fact that the mass of the lower-density plasma in that case diverges to infinity as $n_a \rightarrow 0$.
Hence, similar to treating mass much beyond the peak formation temperature of the analyzed line as hidden, mass much beyond the density inferred from the associated line pair should be considered hidden as well.

\section{Corner Plots of the Fit Parameters to the 2017 Lightcurves}
\label{app:morefits}

\begin{figure}[h!]
\includegraphics{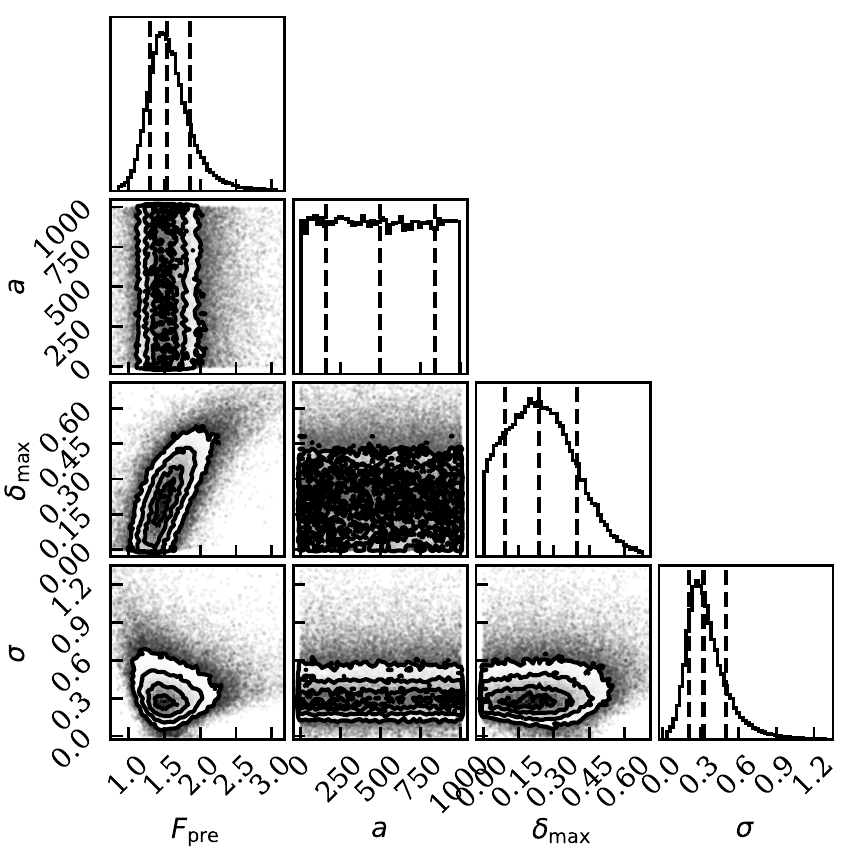}
\caption{March 2017 epoch, MCMC chains from the fit to the \Fexxi\ data in Figure \ref{fig:blc}. \fitunits}
\label{fig:bchainxxi}
\end{figure}

\begin{figure}[h!]
\includegraphics{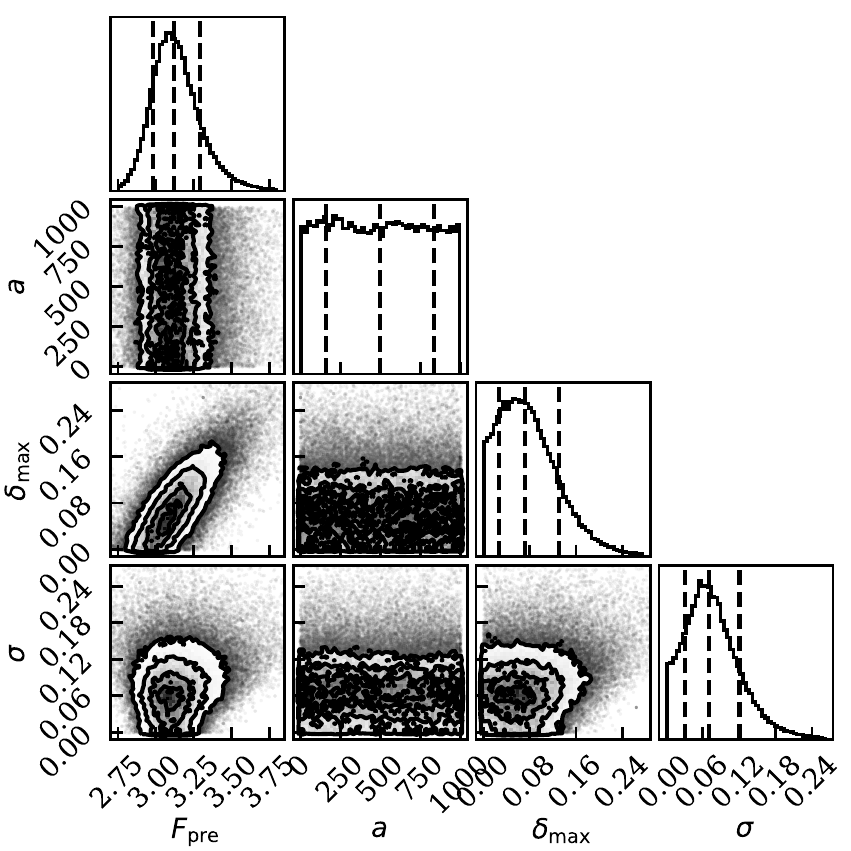}
\caption{March 2017 epoch,  MCMC chains from the fit to the \Fexii\ data in Figure \ref{fig:blc}. \fitunits}
\label{fig:bchainxii}
\end{figure}

\begin{figure}[h!]
\includegraphics{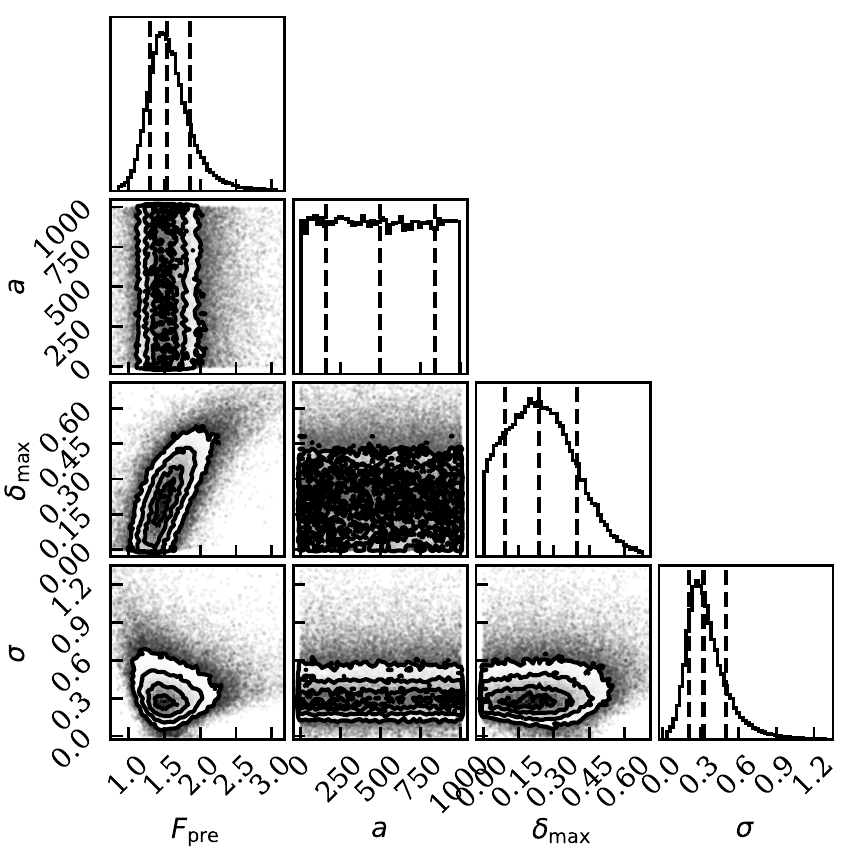}
\caption{August 2017 epoch, MCMC chains from the fit to the \Fexxi\ data in Figure \ref{fig:clc}. \fitunits}
\label{fig:cchainxxi}
\end{figure}

\begin{figure}[h!]
\includegraphics{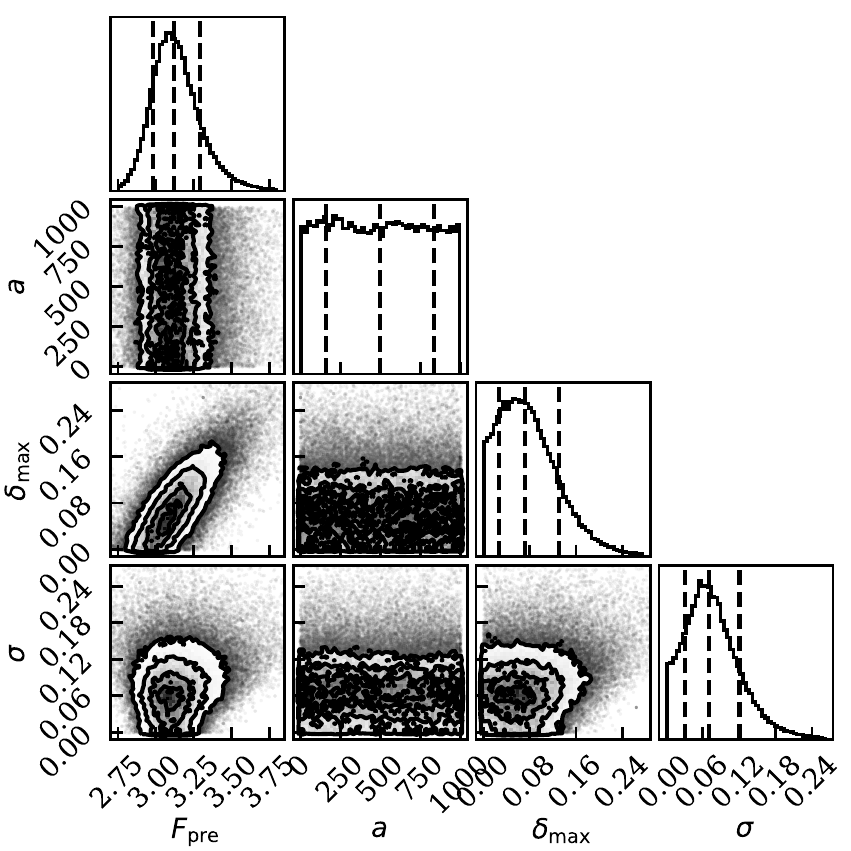}
\caption{August 2017 epoch, MCMC chains from the fit to the \Fexii\ data in Figure \ref{fig:clc}. \fitunits}
\label{fig:cchainxii}
\end{figure}

Figures~\ref{fig:bchainxxi} (\Fexxi) and \ref{fig:bchainxii} (\Fexii) show the posteriors of the fit parameters for the March 2017 observations and Figures~\ref{fig:cchainxxi} (\Fexxi) and \ref{fig:cchainxii} (\Fexii) show the same for the August 2017 observations.

\end{document}